\documentclass[a4paper,11pt]{article}
\usepackage[margin=0.5in]{geometry}
\usepackage{enumerate}   
\usepackage{jcappub}
\usepackage{subcaption}
\usepackage{aas_macros}
\usepackage{multirow}

\usepackage{soul}

\usepackage{amsmath}
\usepackage{amssymb}
\usepackage{bm}
\usepackage{natbib}
\usepackage{xspace}
\usepackage{lipsum}
\usepackage[utf8]{inputenc}
\usepackage[T1]{fontenc}
\usepackage{hyperref}
\usepackage{caption}
\usepackage{verbatim}
\usepackage[dvipsnames]{xcolor}
\usepackage{adjustbox}

\newcommand{\dnull}{{\tt distance-nulling}\xspace}
\newcommand{\knull}{{\tt k-nulling}\xspace}
\newcommand{\planck}{{\sl Planck}\xspace}
\newcommand{\bemu}{{\tt baccoemu}\xspace}
\newcommand{\nmt}{{\tt NaMaster}\xspace}
\newcommand{\Om}{\Omega_{\rm M}}

\hypersetup{
	unicode,
	pdfauthor={Author One, Author Two, Author Three},
	pdftitle={Robust cosmology from cosmic shear with small-scale decontamination},
	pdfsubject={Robust cosmology from cosmic shear with small-scale decontamination},
	pdfproducer={LaTeX},
	pdfcreator={pdflatex}
}
\usepackage{graphicx, color}

\title{\boldmath Robust cosmic shear with small-scale nulling}
\author[a,b]{Giulia Piccirilli,}
\author[c]{Matteo Zennaro,}
\author[c]{Carlos Garc\'ia-Garc\'ia,}
\author[c]{David Alonso}

\affiliation[a]{Dipartimento di Fisica, Università degli Studi di Torino, via P. Giuria 1, 10125 Torino, Italy}
\affiliation[b]{INFN – Istituto Nazionale di Fisica Nucleare, Sezione di Torino, via P. Giuria 1, 10125 Torino, Italy}
\affiliation[c]{Department of Physics, University of Oxford, Denys Wilkinson Building, Keble Road, Oxford OX1 3RH, United Kingdom}

\emailAdd{giulia.piccirilli@unito.it}

\abstract{Standard cosmological weak lensing analyses using cosmic shear are inevitably sensitive to small-scale, non-linear clustering from low-redshift structures. The need to adequately model the clustering of matter on this non-linear regime, accounting for both gravitational and baryonic effects, adds significant uncertainty to weak lensing studies, particularly in the context of near-future Stage-IV datasets. In this paper, inspired by previous work on so-called ``nulling'' techniques, we present a general method that selects the linear combinations of a given tomographic cosmic shear dataset that are least sensitive to small-scale non-linearities, by essentially suppressing the contribution from low-redshift structures. We apply this method to the latest public cosmic shear data from the Dark Energy Survey, DES-Y3, that corresponds to 3 years of observation, and show: a) that a large fraction of the signal is dominated by the single mode that is most affected by non-linear scales, and b) that removing this mode leads to a $\sim1\sigma$ upwards shift in the preferred value of $S_8\equiv\sigma_8\sqrt{\Om/0.3}$, alleviating the tension with current CMB data. However, the removal of the most contaminated mode also results in a significant increase in the statistical uncertainties. Taking this into account, we find this shift to be compatible with a random fluctuation caused by removing this most-contaminated mode at the $\sim1.4\sigma$ level. We also show that this technique may be used by future Stage-IV surveys to mitigate the sensitivity of the final constraints to baryonic effects, trading precision for robustness.}
\begin{document}
\maketitle
\flushbottom

\section{Introduction}
  Weak gravitational lensing, the deflection of photon trajectories from background sources caused by the gravitational potential of the foreground large-scale structure, is a key cosmological probe \cite{astro-ph/9912508,1710.03235,2501.07938}. Imaging galaxy surveys are highly sensitive to weak lensing through the correlated distortion it causes to the shapes of background galaxies, the so-called ``cosmic shear'' effect. Since its first detection \cite{1990ApJ...349L...1T,astro-ph/0002500,astro-ph/0003008,astro-ph/0003014,astro-ph/0003338}, cosmic shear surveys have grown in area and depth, and are now able to place highly stringent constraints on the cosmological parameters that govern the evolution of structure at late times.

  The latest generation of cosmic shear experiments, consisting of the Dark Energy Survey (DES) \cite{2105.13544}, the Hyper Suprime-Cam survey (HSC) \cite{2304.00701}, and the Kilo-Degree Survey (KiDS) \cite{2007.15633}, have been able to measure the ``clumping'' parameter $S_8$ with a precision of a few percent ($\sigma(S_8)\sim0.02$). Commonly defined as $S_8\equiv\sigma_8\sqrt{\Om/0.3}$, where $\Om$ is the non-relativistic matter energy fraction, and $\sigma_8$ is the standard deviation of the linear overdensities in spheres of $8\,h^{-1}{\rm Mpc}$ radius, the clumping parameter characterises the level of inhomogeneity in the matter distribution at late times. Interestingly, current measurements of $S_8$ seem to consistently lie below the value inferred from observations of the Cosmic Microwave Background (CMB) at early times, and extrapolated to $z=0$ assuming a $\Lambda$CDM cosmology \cite{1807.06209,2203.06142,2105.12108,2305.17173,2403.13794}. Evidence for a low value of $S_8$ has also been found in other data combinations, including cross-correlations between weak lensing measurements and the galaxy overdensity \cite{2105.03421,2111.09898,2204.10392}, the abundance of massive clusters detected via the thermal Sunyaev-Zel'dovich (tSZ) effect \cite{1502.01597} (although see more recent analyses e.g. \cite{1904.07887,2401.02075,2402.08458}), and the spectrum of the tSZ fluctuations themselves \cite{1712.00788}. However, the highest-significance evidence of a low $S_8$ seems to originate from cosmic shear data \cite{2105.12108}, which is a direct probe of clustering in the late Universe ($z\lesssim0.5$), almost free from uncertainties governing the detailed clustering of galaxies. It is therefore important to dissect current cosmic shear measurements, to understand what elements of the data show the strongest evidence of tension with the CMB.

  Along these lines, it has been recently argued that the origin of the $S_8$ tension could lie in the presence of strong baryonic feedback effects: highly-energetic outflows driven by supermassive black holes at the centres of galaxies can eject a large fraction of the gas content beyond the halo virial radius, thus suppressing the clustering of matter on small scales \cite{2023MNRAS.525.5554P,2024MNRAS.534..655B,2024arXiv241019905M}. If this effect is ignored or underestimated, it leads to a low inferred value of $S_8$ \cite{2206.11794, 2303.05537, 2023MNRAS.525.5554P, 2305.17173, 2024MNRAS.534..655B}. The analysis of some probes of the cosmic gas, including the kinematic Sunyaev-Zel'dovich effect \cite{2024MNRAS.534..655B}, and X-ray observations \cite{2309.11129,2412.12081}, has shown some evidence that the strength of baryonic feedback may be larger than predicted by many hydrodynamical simulations, although this is not entirely clear \cite{2110.02228,2301.01354,2309.02920,2403.13794,2403.20323,2502.06687}, and it may not be enough to explain the $S_8$ value measured from other observables \cite{2309.07959}. It has also been shown that, irrespective of the impact of baryonic feedback, an inaccurate modelling of the non-linear matter power spectrum driven by gravity alone can also lead to a significant mis-estimation of $S_8$ \cite{2303.05537,2410.22272}.
  
  In general, the impact of small-scale, non-linear effects in cosmic shear observations must be carefully scrutinised as a potential source of theoretical uncertainties.
  This is particularly important for cosmic shear: since the weak lensing of sources at a given redshift is caused by all of the intervening large-scale structure, the cosmic shear from source galaxies at virtually any redshift always receives a significant contribution from structures at very low redshifts which, although physically small in size, subtend relatively large angular scales. Isolating the contribution from these structures would allow us to explore whether the $S_8$ tension is indeed driven by small-scale effects at late times. This can in principle be achieved by combining cosmic shear data from galaxies at different redshifts. Schematically, the cosmic shear of a  given low-redshift source sample is caused by some of the same structure causing the shear of a higher-redshift sample, and therefore the former can be used to ``clean'' the latter from the impact of this low-redshift contribution. This is at the heart of several so-called ``nulling'' approaches proposed in the literature \cite{astro-ph/0501451,1312.0430,1809.03515,2007.00675,2009.01792,2012.04672,2502.02246}. In essence, nulling methods perform a linear transformation to the cosmic shear data from a set of tomographic redshift bins to isolate the contributions from well-defined redshift ranges. Different approaches have been used to achieve this. The most common approach studied in the literature is the so-called ``Bernardeau-Nishimichi-Taruya (BNT)'' transformation, first proposed by \cite{1312.0430}. This linear transformation aims to produce modified kernels that are compact and as disjoint as possible in redshift, making it possible to identify angular scale cuts that are more easily associated to well-motivated physical scale cuts. The method has been exploited to quantify the impact of small-scale systematics in cosmic shear data, both in real and Fourier space \citep{1809.03515,2007.00675,2012.04672,2107.10277}.

  In this paper we will use an alternative approach, with the same motivation in mind, closer in spirit to the method outlined in \cite{astro-ph/0501451}. Rather than maximising the compactness of the resulting radial kernels, we will instead use a minimisation framework to rank the different orthogonal modes of the tomographic data vector by their sensitivity to the small-scale power, and project out the most contaminated modes. As we will show, mathematically the method is equivalent to the nulling techniques recently proposed in the context of lensing from the CMB and line intensity mapping \citep{1909.02615, 2011.06582, 2106.09005, 2208.04253}, in which tracers at different redshifts are combined to minimise the contribution from low-redshift structures to a given lensing map. The method is also closely related to general 3D data compression techniques proposed in the literature \cite{1707.08950,1903.04957}. We will apply this methodology to state-of-the-art weak lensing data from the Dark Energy Survey (DES) Year 3 data release \cite{2011.03407}, and study the potential impact of small, non-linear scales on the current $S_8$ constraints. We will also quantify the potential uses of this approach in the analysis of future Stage-IV surveys, such as the Rubin Observatory Legacy Survey of Space and Time (LSST) \cite{0805.2366,1809.01669}, particularly in the context of baryonic effects.

  The paper is structured as follows. Section~\ref{sec:method} describes the theoretical background behind cosmic shear observations, including the most relevant systematic uncertainties caused by complex small-scale physics, and presents the specific nulling technique proposed here in detail. Section~\ref{sec:data} describes the DES cosmic shear data and the methods used to estimate and analyse their tomographic two-point statistics. The results obtained applying our nulling approach to the DES data, and to a synthetic LSST-like data vector, are presented in Section~\ref{sec:res}, with an emphasis on the impact of small-scale effects on the measurements of $S_8$. We then conclude in Section~\ref{sec:concl}.

\section{Methodology}\label{sec:method}
  \subsection{Theory background}\label{ssec:method.theory}
    Consider measurements of the cosmic-shear signal in $N_t$ tomographic bins with normalised redshift distributions $p_i(z),\,\,i\in[1,N_t]$. Assuming a flat $\Lambda$CDM cosmology, the angular power spectrum of the cosmic shear $E$-modes of any pair of these bins is given by
    \begin{equation}
      \label{eq:theory_ps}
      C_{\ell}^{ij} = G^2_{\ell}\int \frac{d\chi}{\chi^2} q_i(\chi)q_j(\chi)P\left(k = \frac{\ell+1/2}{\chi}, z(\chi)\right).
    \end{equation}
    Here $P(k,z)$ is the non-linear matter power spectrum at redshift $z$ and wavenumber $k$, and $\chi$ is the radial comoving distance. The radial lensing kernel of bin $i$ is related to the redshift distribution via
    \begin{equation}
      \label{eq:lensing_kernel}
      q_i(\chi)\equiv \frac{3H_0^2\Om}{2a(\chi)c}\,\chi\int_{z(\chi)}^\infty\,dz'\,p_i(z')\left(1-\frac{\chi}{\chi(z')}\right),
    \end{equation}
    where $H_0 $ is the Hubble parameter, and $a$ is the scale factor. Equation \ref{eq:theory_ps} holds in the Limber approximation \citep{Limber_1953}, which assumes that the radial kernels involved are much broader than the distance over which matter fluctuations show significant correlation. Finally, the prefactor $G_{\ell}\equiv\sqrt{(\ell+2)!/(\ell -2)!}/(\ell+1/2)^2$ accounts for the difference between the 3D Laplacian of the gravitational potential and the angular Hessian of the corresponding lensing potential \cite{Kilbinger_2017}, and is negligibly close to 1 beyond $\ell\sim10$.

    An important source of astrophysical systematics in weak lensing measurements is the presence of intrinsic alignments (IA), i.e., the coherent alignment in the intrinsic shape of nearby galaxies. The simplest model to describe IA is the so-called Non Linear Alignment model (NLA, \cite{Hirata_2004}), in which intrinsic galaxy shapes are proportional to the local tidal field. In this case, IA can be simply described by adding a local contribution to the lensing kernel with the form
    \begin{equation}
    \label{eq:ia_model}
      q^{\rm IA}_i(\chi) = -A_{\rm IA}(z)p_i(z)\frac{dz}{d\chi},
    \end{equation}
    where $A_{\rm IA}$ is the alignment amplitude, with a redshift evolution that is commonly parametrised as \cite{Abbott_2017}
    \begin{equation}
      A_{\rm IA}(z) = A_{\rm IA,0}\left(\frac{1+z}{1+z_0}\right)^{\eta_{\rm IA}}\frac{0.0139\Om}{D(z)}.
    \end{equation}
    Here $D(z)$ is the linear growth factor, and $z_0=0.65$. The amplitude at $z=0$, $A_{\rm IA,0}$, and the evolution parameter $\eta_{\rm IA}$ are free parameters of the model that must be constrained by the data.

  \subsection{Small-scale contaminants in weak lensing}\label{ssec:method.bar_effect}
    Equation \ref{eq:theory_ps} shows how the shear angular power spectrum receives contributions from the underlying matter power spectrum at scales, in principle, infinitely small. In practice, the lensing kernels select the relative contribution of the different scales. However, for current galaxy surveys, scales as small as $k=5 ~ h ~ \mathrm{Mpc}^{-1}$ have a significant contribution \citep[see, e.g.,][]{2403.13794}, and even smaller scales will be probed by upcoming surveys. Modelling these scales is not straightforward for a number of reasons. First, perturbative approaches used to include nonlinearities to the linear matter power spectrum typically break down at $k \approx 0.2 ~ h ~ \mathrm{Mpc}^{-1}$ \citep{CrocceScoccimarro2006}; N-body simulations can predict the nonlinear matter power spectrum to smaller scales, but come at a prohibitively expensive computational cost, thus requiring the use of machine learning-based emulators \citep{LawrenceEtal2017,KnabenhansEtal2019,AnguloEtal2021,KnabenhansEtal2021,MoranEtal2023,DeRoseEtal2023}. Second, the growth of matter perturbation on small scales is strongly affected by baryonic effects,  i.e. astrophysical processes that redistribute the baryonic mass contained in haloes, such as black hole accretion and feedback, star and galaxy formation, stellar winds and supernova explosions. The details of these processes are in principle unknown, to the point that different hydrodynamical simulations that try to simultaneously evolve dark matter perturbations and astrophysical processes give very different predictions of the effects baryons have on the matter power spectrum -- so different that they are incompatible among each other \citep{1104.1174,1105.1075,1603.03328,1702.02064,1707.03397,1801.08559,1906.00968}. Finally, even if we were able to perfectly model the matter power spectrum at small scales, the lensing angular power spectrum is contaminated by the intrinsic alignment of galaxy shapes induced by the large-scale structure of the Universe; this signal is mixed with the desired weak lensing signal, but the details of its origin are not clear. Developing a first-principles model for IA is therefore complicated, although the requirements for its accuracy are likely less stringent than those for the small-scale matter power spectrum \cite{2311.16812}.

    In this work, we use the matter power spectrum emulator \bemu\footnote{\url{https://baccoemu.readthedocs.io/en/latest/}.} to obtain predictions down to $k_{\rm max} = 5 ~ h ~ \mathrm{Mpc}^{-1}$.  This version of \bemu includes an extension of the parameter space \citep[presented in ][]{2303.05537} in which the expected accuracy on the matter power spectrum can reach $\approx 3\%$ at $k=5 ~ h ~ \mathrm{Mpc}^{-1}$. This level of accuracy was shown to be enough for current weak lensing data \citep{2303.05537,2403.13794}. However, we still have to deal with the small-scale contamination induced by baryonic effects. In terms of the matter power spectrum, this is typically quantified as the ratio between a power spectrum obtained accounting for gravity and hydrodynamical processes (i.e. a full-physics description of the growth of structures), $P_{\rm full}$, and its corresponding counterpart in which only the gravitational growth is accounted for, $P_{\rm GrO}$,
    \begin{equation}
      S(k) = \dfrac{P_{\rm full}}{P_{\rm GrO}}.
    \end{equation}
    We will refer to $S(k)$ as the scale-dependent ``baryonic boost factor''. Different models exist for $S(k)$. They include modifications of the halo model \citep[e.g.][]{1105.1075,SemboloniEtal2013,MohammedEtal2014,Fedeli2014,MeadEtal2015,DebackereEtal2020,MeadEtal2021}, approaches based on principal component analyses \citep[e.g.][]{EiflerEtal2015,HuangEtal2019}, corrections applied through machine learning tools \citep[e.g.][]{TrosterEtal2019,Villaescusa-NavarroEtal2020}, or applying Effective Field Theory \citep[][]{BragancaEtal2021}, or even directly interpolating
 between hydrodynamical simulation outputs through an emulator \citep{SchallerEtal2024b}. On the other hand, this effect can be modelled by employing physically motivated displacements of particles belonging to each halo in $N$-body simulations, an approach called \textit{baryonification} or \textit{Baryon Correction Model} (BCM) \citep{SchneiderTeyssier2015,SchneiderEtal2019,AricoEtal2020, 2104.04165}. In this work, when creating synthetic data vectors that include the effects of baryons, we will use the baryonic boost emulator included in the \bemu package \citep{AnguloEtal2021,AricoEtal2021b}. This is based on the BCM applied to dark matter-only simulations and has an accuracy better than 1-2\% up to $k=5~h~\mathrm{Mpc}^{-1}$. It has 7 free parameters: $M_{\rm c}$ controls the typical halo mass at which haloes have lost half of their gas component, $\beta$ is the slope of the dependence of the gas profile on halo mass, $M_{\rm inn}$, $\vartheta_{\rm inn}$, and $\vartheta_{\rm out}$ define a broken power law describing the virialised gas profile, $\eta$ controls the maximum distance reached by the gas ejected from haloes, and $M_{1,z_0,\mathrm{cen}}$ controls the typical mass of central galaxies at redshift $z=0$. These parameters in principle evolve with redshift and, therefore, in the absence of a a clear redshift evolution model, should be left free to vary in each different redshift bin. However, for the sake of simplicity, we will assume the same baryonic parameters at all redshifts, without losing generality for our results. We contaminate our synthetic data vector by obtaining, with \bemu, a matter power spectrum suppression $S(k)$ that reproduces the one found in the BAHAMAS hydrodynamical simulation \citep{McCarthyEtal2017}. To do so, we use $\log_{10}[M_{\rm c}/(h^{-1} \mathrm{M}_{\odot})] = 13.58$, $\log_{10}[\eta] = -0.27$, $\log_{10}[\beta] = -0.33$, $\log_{10}[M_{1, z_0, \mathrm{cen}}/(h^{-1} \mathrm{M}_{\odot})] = 12.04$, $\log_{10}[\vartheta_{\rm out}] = 0.25$, $\log_{10}[\vartheta_{\rm inn}] = -0.86$, and $\log_{10}[M_{\rm inn}/(h^{-1} \mathrm{M}_{\odot})] = 13.4$ \citep{AricoEtal2020}.

  \subsection{Nulling transformation and robust lensing}\label{ssec:method.null}
    In this section, we introduce the nulling transformation used in this analysis. In particular, our goal is to isolate the contributions to the cosmic shear signal that come from a specific low-redshift range or, equivalently, to suppress the contribution from small-scale fluctuations at very low redshifts.

    \subsubsection{A toy example}\label{sssec:method.null.toy}
      \begin{figure}[h]
          \centering
          \includegraphics[width=\textwidth]{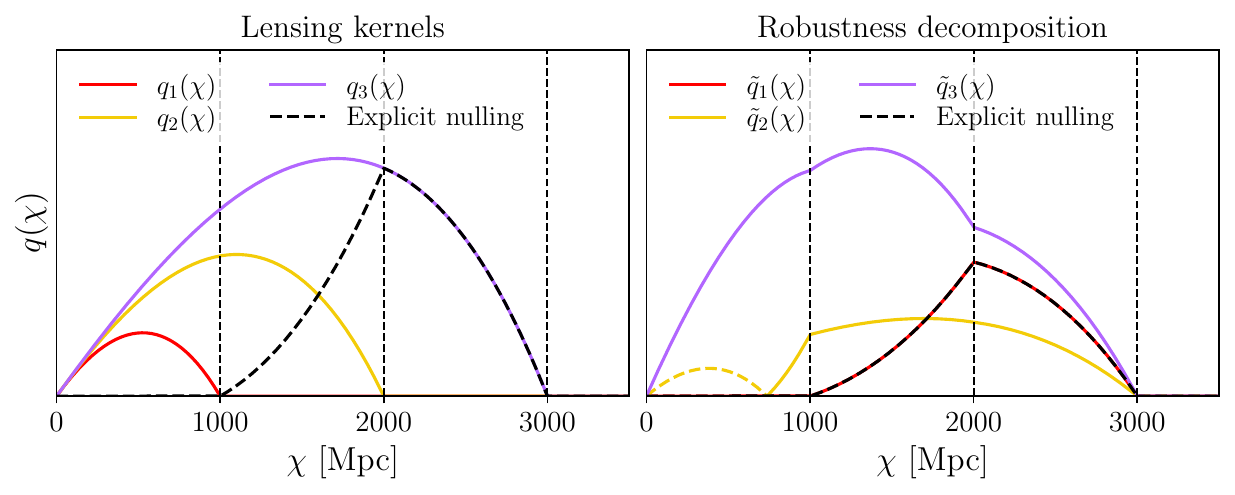}
          \caption{{\sl Left:} lensing kernels associated with three source planes at $\chi=1000$, 2000, and 3000 Mpc (red, yellow, purple), and the effective kernel constructed by explicitly nulling (Equation~\ref{eq:explicit_nulling}) the contribution of all structure below $\chi=1000$ Mpc (dashed black). {\sl Right:} effective kernels found using the \dnull transform described in Section \ref{sssec:method.null.chi} for the same set of source planes. The first set of ``eigenweights'' (red) are equivalent to the explicit nulling shown in the left panel.}
          \label{fig:toy_example}
      \end{figure}
      Let us start by illustrating the logic behind the nulling transform with an idealised scenario (see \cite{2106.09005}). Consider the case of three tomographic redshift bins with delta-like redshift distributions, at distances $\chi_i$, with $i\in\{1,2,3\}$. The radial kernel for the $i$-th bin is simply
      \begin{equation}
        q_i(\chi)=Q_L(\chi)\,\left(1-\frac{\chi}{\chi_i}\right)\Theta(\chi<\chi_i),\hspace{12pt}Q_L(\chi)\equiv\frac{3H_0^2\Om}{2a(\chi)c}\chi,
      \end{equation}
      where we have included the Heavyside function $\Theta$ to make it explicit that the kernel is exactly zero for $\chi>\chi_i$. Ignoring the prefactor $Q_L(\chi)$, common to all lensing kernels, the kernels in this case are simple linear functions of $\chi$, defined by a single coefficient ($1/\chi_i$). We can thus construct a linear combination of the two higher-redshift bins with an effective kernel that is exactly the same as that of the first bin for $\chi<\chi_1$. Subtracting the first bin from this combination, the result is a linear combination of the bins in which the kernel at distances $\chi<\chi_1$ is exactly zero. Explicitly, consider a linear combination of the last two bins, with linear coefficients $\{a_2,a_3\}$. To match the kernel of the first bin, we require
      \begin{equation}
        a_2(1-\chi/\chi_2)+a_3(1-\chi/\chi_3)=1-\chi/\chi_1.
      \end{equation}
      Matching coefficients on both sides, we obtain the solution
      \begin{equation}
        a_2=\frac{1/\chi_1-1/\chi_3}{1/\chi_2-1/\chi_3}, \hspace{12pt}
        a_3=\frac{1/\chi_2-1/\chi_1}{1/\chi_2-1/\chi_3}.
      \end{equation}
      Subtracting the first bin from this combination, we find that the following combination of kernels will be exactly zero for $\chi<\chi_1$\footnote{Note that we applied an additional normalisation factor to the final linear combination to make the functional form of the weight coefficients $w_i$ more appealing.}
      \begin{equation}\label{eq:explicit_nulling}
        \tilde{q}_w(\chi)\equiv \sum_{i=1}^3w^i\,q_i(\chi),\hspace{12pt}
        w^1=1-\frac{\chi_2}{\chi_3},\hspace{6pt}w^2=\frac{\chi_2}{\chi_3}-\frac{\chi_2}{\chi_1},\hspace{6pt}
        w^3=\frac{\chi_2}{\chi_1}-1.
      \end{equation}

      The left panel of Fig. \ref{fig:toy_example} shows the lensing kernels for $(\chi_1,\,\chi_2,\,\chi_3)=(1000,\,2000,\,3000)$ Mpc, together with the effective kernel for the null combination above which, as expected, is non-zero only for $\chi>\chi_1=1000\,{\rm Mpc}$.

    \subsubsection{General low-redshift nulling: \dnull case}\label{sssec:method.null.chi}
      Let us now consider the more general problem of splitting our data vector into linear combinations ranked by the contribution to them of structures at low redshifts. Consider one such combination, characterised by a set of weights ${\bf w}\equiv(w^1,\cdots,w^{N_t})$, where $N_t$ is the number of tomographic redshift bins. The radial kernel associated with this combination is $\tilde{q}_w(\chi)=\sum_i w^i\,q_i(\chi)$. 

      As a measure of how much support $\tilde{q}_w(\chi)$ has at low redshifts, let us consider the following quantity:
      \begin{equation}\label{eq:R}
        {\cal R}_w\equiv \int_0^{\chi_L}d\chi\,\tilde{q}_w^2(\chi)={\bf w}^T{\sf R}{\bf w},
      \end{equation}
      where $\chi_L$ is the comoving distance to some low redshift, and we have defined the matrix
      \begin{equation}
        {\sf R}_{ij}\equiv \int_0^{\chi_L}d\chi\,q_i(\chi)q_j(\chi).
      \end{equation}
      It is important to note that ${\sf R}$ is a symmetric, positive-definite matrix, and, therefore,  ${\cal R}_w$ is a positive function of the weights bounded from below by 0. We would like to minimise ${\cal R}_w$, while avoiding the trivial solution ${\bf w}=0$. For this, let us impose a normalisation constraint on ${\bf w}$, for example making it unit-normed, via a Lagrange multiplier. We will then consider minimising the quantity
      \begin{equation}
        {\cal L}({\bf w})\equiv {\cal R}_w+\lambda(1-{\bf w}^T{\bf w}).
      \end{equation}
      Setting the gradient of ${\cal L}({\bf w})$ to zero, we obtain the eigenvalue equation
      \begin{equation}
        {\sf R}{\bf w}=\lambda{\bf w},
      \end{equation}
      where the weight vectors we are looking for are the eigenvectors {\bf w} of the matrix ${\sf R}$, with eigenvalues given by the Lagrange multiplier $\lambda$. Furthermore, for any eigenvector ${\bf w}_a$, the corresponding value of ${\cal R}_{w_a}$ is given by its associated eigenvalue $\lambda_a$: ${\cal R}_{w_a}={\bf w}^T_a{\sf R}{\bf w}_a=\lambda_a{\bf w}_a^T{\bf w}_a=\lambda_a$, since ${\bf w}^T_a{\bf w}_b=\delta_{ab}$ by construction. These ``eigenweights'' can be ranked in order of increasing $\lambda_a$, in which case, the first weight vector ${\bf w}_a$ corresponds to the most ``robust'' combination of our data (i.e. the one with the lowest possible contribution from structures at low redshifts), and all subsequent weight vectors are increasingly less robust.
      
      This simple procedure depends only on the value of the minimum distance $\chi_L$, and yields scale-independent weight vectors. We will refer to it in what follows as \dnull. Once the eigenweights have been found, the angular power spectra of their associated weighted shear maps are related to the full set of tomographic power spectra via
      \begin{equation}\label{eq:cl2cl}
        \tilde{C}^{ab}_\ell=\sum_{ij}w_a^iw_b^jC^{ij}_\ell.
      \end{equation}
      Note that the indices $\{a,b\}$ run over the level of contamination from small-scales, whereas the indices $\{i, j\}$ run over the elements of the weight vectors ${\bf w}_a$. Therefore, we can then consider including only a limited number of eigenmodes, for example taking only the first few most robust modes (e.g $\tilde{C}_{\ell}^{11}$), or the most contaminated ones (e.g. $\tilde{C}_{\ell}^{N_t N_t}$). An interesting feature of Equation \ref{eq:cl2cl} is that the same weighting scheme can be applied to the data and the theoretical predictions, reducing the modifications needed to existing analysis pipelines to a minimum.

      Applying this formalism to the toy example described in the previous section (tomographic bins at three discrete redshifts), with $\chi_L=\chi_1$, leads to three modified radial kernels defined by each of the three eigenweights: $\tilde{q}_a(\chi)\equiv \sum_iw_a^i\,q_i(\chi)$. These are shown in the right panel of Figure~\ref{fig:toy_example}. We can see that the first kernel, associated with the most robust mode, is identical to the nulled kernel found explicitly in the previous section (up to an irrelevant normalisation factor). This is not surprising: we found that this particular linear combination achieves ${\cal R}_w=0$, which is the lowest possible value this quantity can take, and must therefore correspond to the lowest eigenvalue of ${\sf R}$. Interestingly, we find that the most contaminated mode has an associated kernel that has a significantly higher amplitude than the first two modes. Since the weight vectors are normalised to unit norm, this implies that a large fraction of the cosmic shear signal in our data is concentrated in this mode. We will find this result consistently in all other cases explored here, and it is a consequence of the dominant role that small-scale, low-redshift structures play in cosmic shear observations. Minimising the dependence on these non-linear scales will invariably come at the cost of discarding a large fraction of the signal.

    \subsubsection{General small-scale  nulling: {\tt k-nulling} case}\label{sssec:method.null.k}
      \begin{figure}[h]
          \centering
          \includegraphics[width=\textwidth]{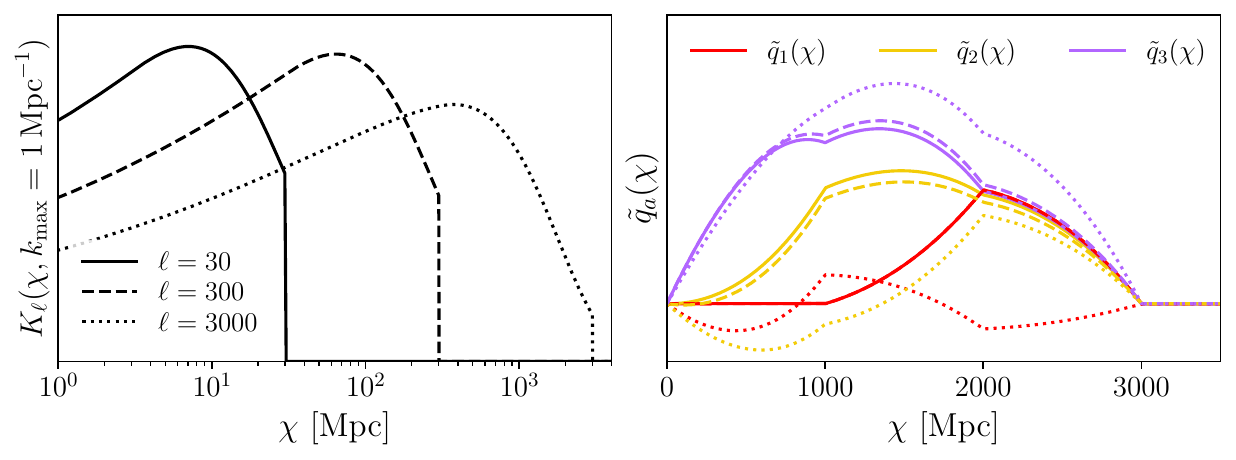}
          \caption{{\sl Left:} scale-dependent kernel in the \knull transform described in Section \ref{sssec:method.null.k}, for $k_{\rm max}=1\,{\rm Mpc}^{-1}$ and different values of $\ell$. {\sl Right:} resulting rotated radial kernels. Note that $\tilde q_1$ is the same for $\ell\lesssim k_{\rm max} \chi_1 = 1000$, where $\chi_1$ is the distance to the first redshift bin.}
          \label{fig:toy_example_elldep}
      \end{figure}
      The robustness metric introduced in Equation \ref{eq:R} may be generalised as
      \begin{equation}\label{eq:R-knull}
        {\cal R}_w\equiv \int d\chi\,K(\chi)\,\tilde{q}_w^2(\chi)=w_iw_j\int d\chi\,K(\chi)\,q_i(\chi)q_j(\chi)\equiv{\bf w}^T{\sf R}{\bf w},
      \end{equation}
      where $K(\chi)$ is a general kernel with support mostly at low redshifts, which can be used to connect ${\cal R}_w$ more closely with quantities related to the shear power spectrum we wish to keep under control. For instance, in the Limber approximation, we can write the contribution to the $C_\ell$ from physical scales above a given 3D wavenumber $k_{\rm max}$ as
      \begin{equation}
        \frac{\ell^2}{2\pi}C_\ell^{\rm SS}\equiv\int_0^{\ell/k_{\rm max}}\frac{d\chi}{2\pi}\,\tilde{q}_w^2(\chi)\frac{\ell^2}{\chi^2}P(\ell/\chi,z(\chi)),
      \end{equation}
      where ${\rm SS}$ stands for ``small scales''.      Thus, a well-motivated kernel is
      \begin{equation}
        K_\ell(\chi,k_{\rm max})=\left[k^2P(k)\,\Theta(k>k_{\rm max})\right]_{k=\ell/\chi}.
      \end{equation}
      In this case, the resulting eigenweights are scale-dependent: the contributions from different redshift bins to the weighted shear map associated with one of these weight vectors varies for different angular multipoles. The $\ell$-dependence of the weights must therefore be taken into account when constructing the angular power spectra of these weighted maps from the original set of tomographic power spectra (Equation~\ref{eq:cl2cl}). In what follows we will refer to this transformation as \knull. This approach is similar in spirit to the $k$-cut method of \cite{2107.10277}, which instead exploits the BNT transform.

      The left panel of Figure~\ref{fig:toy_example_elldep} shows the form of $K(\chi)$ for different $\ell$s and a scale cut of $k_{\rm max}=1\,{\rm Mpc}^{-1}$. As the plot shows, any attempt to minimise the sensitivity to these physical scales at $\ell\gtrsim3000$ would involve suppressing structures at distances below $\chi\sim3000\,{\rm Mpc}$. The radial kernels associated with the corresponding eigenweights are shown in the right panel of the figure. We can see that, as we move to smaller angular scales, more power is transferred from the first two eigenmodes to the third one (i.e. the most contaminated one), since suppressing the small-scale dependence becomes more difficult. It is interesting to note that the first eigenmode is exactly the same for both $\ell=30$ and $\ell=300$, and in fact it is easy to see that this is the case for all $\ell\lesssim k_{\rm max}\chi_1=1000$, where $\chi_1$ is the distance to the first redshift bin (which is a $\delta$-function in this example). 

\section{Data analysis}\label{sec:data}
  This section describes briefly the dataset used in this analysis, the methods used to measure cosmic shear power spectra, and the likelihood used to extract cosmological constraints from these measurements.

  \subsection{Cosmic shear power spectra from  DES-Y3}\label{ssec:data.descls}
    We use the three-years public data release of the Dark Energy Survey (DES-Y3\footnote{\url{https://des.ncsa.illinois.edu/releases/y3a2/Y3key-catalogs}}). The sample used, and the methods employed to extract power spectrum measurements were described in detail in \cite{2403.13794}, and we only provide a short summary here for completeness.
    
    The DES-Y3 dataset includes $\sim100$ million galaxies over 5000 deg$^2$, for an effective number density of $n_{\rm gal} = 5.59~{\rm gal/arcmin}^2$. The sample was divided into four broad tomographic redshift bins in the photometric redshift range $z_{\rm ph} = [0, 1.5]$, each having approximately the same number density of sources. Galaxy shapes were measured via {\tt METACALIBRATION} \cite{Huff_2017}. As in \cite{Gatti_2021}, in each tomographic bin we first subtract the mean ellipticity of all sources and correct for the average multiplicative bias of the sample, $m$, estimated through the response tensor $R$ (not to be confused with the nulling metric ${\cal R}_w$ and associated matrix ${\sf R}$, introduced in Section \ref{ssec:method.null}).
    
    \begin{figure}[h!]
      \centering
      \includegraphics[width = 0.9\linewidth]{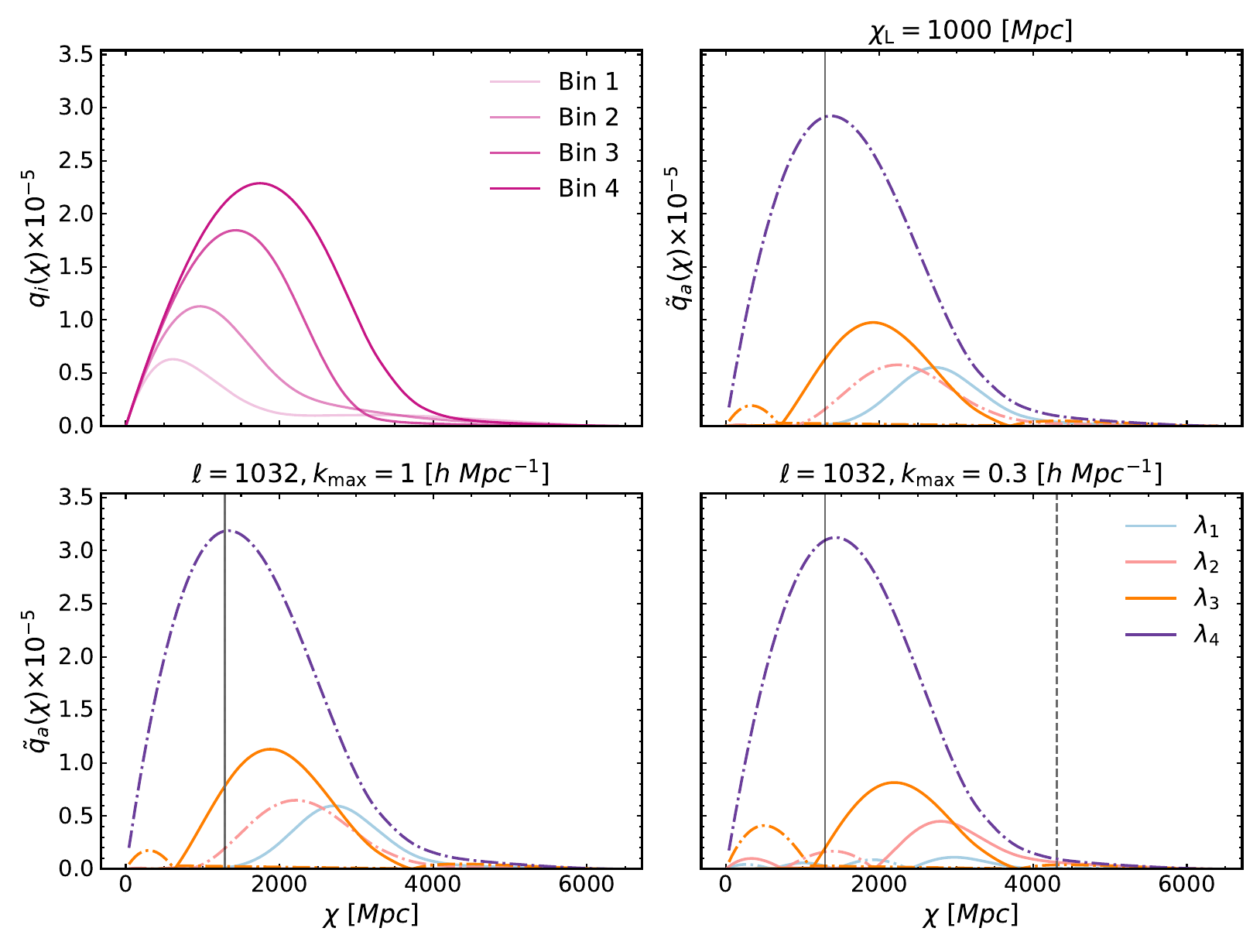}
      \caption{DES-Y3 lensing kernels as a function of comoving distance. {\sl Top left:} input DES-Y3 kernels for the four considered bins. {\sl Top right:} ``rotated'' version for the \dnull case using $\chi_L = 100~{\rm Mpc}$. {\sl Bottom panels:} lensing kernels rotated according to the \knull transform, shown for two different values of $k_{\rm max}$. Negative values are shown as dotted-dashed lines while the black vertical line marks the comoving distance to the mean redshift of the first tomographic bin.}
      \label{fig:qi_qw_desy}
    \end{figure}
    The lensing kernels, for these tomographic samples, defined in Equation \ref{eq:lensing_kernel}, are displayed in the top left panel of Figure \ref{fig:qi_qw_desy}. 
    The radial kernels associated with the different eigenmodes in the \dnull are shown in the top-right panel for $\chi_L = 1000~{\rm Mpc}$, with the \knull eigenmode kernels for $k_{\rm max}=1\,h~{\rm Mpc}^{-1}$ and $k_{\rm max}=0.3\,h~{\rm Mpc}^{-1}$ shown in the bottom left and right panels, respectively, for $\ell=1032$. 
    The vertical lines in the figure mark the positions of the mean redshifts of the four redshift bins. We can see that the first eigenmode in the \dnull case is effectively insensitive to all structure below the distance to the first redshift bin of the sample. This is also the case for \knull if $k_{\rm max}=1\,h~{\rm Mpc}^{-1}$ at $\ell\sim1000$. However, trying to suppress the small-scale structure with a more conservative $k_{\rm max}=0.3\,h~{\rm Mpc}^{-1}$ is not possible as the amplitude of the first kernel becomes significantly suppressed.
   
    The cosmic shear power spectra were measured using the pseudo-$C_\ell$ formalism \cite{Hivon_2002}, as implemented in \nmt\footnote{\url{https://namaster.readthedocs.io/en/latest/}} \cite{Alonso_2019}. In particular, when generating the cosmic shear maps and their associated masks, as well as estimating the impact of noise auto-correlation bias, we followed the methodology described in  \cite{Nicola_2020}. Specifically, the shear field in pixel $p$ is estimated by averaging over the ellipticities of all galaxies in the pixel
    \begin{equation}
      \gamma^\alpha_p = \frac{\sum_{n\in p} w_n\,e^\alpha_n}{\sum_{n\in p}w_n},
    \end{equation}
    where $w_n$ represents the shape measurement weight of the $n$-th galaxy in the pixel, and $e^\alpha_n$ is its $\alpha$-th ellipticity component ($\alpha\in\{1,2\}$). The mask was then selected to be proportional to the sum of weights in each pixel, $m_p\propto \sum_{n\in p}w_n$, which approximately tracks the inverse variance of the shear measurement. All maps were generated using a {\tt HEALPix}\footnote{\url{https://healpix.sourceforge.io/}} \cite{Gorski_2004} pixelisation scheme with a resolution parameter $N_{\rm side} = 4096$.

    The power spectra were estimated in bandpowers as defined in \cite{2403.13794}, using only those in the scale range $30<\ell<2000$. As discussed in \cite{Nicola_2020}, the noise bias was estimated analytically and subtracted from the measured spectra. The Gaussian part of the covariance matrix for the measured spectra was computed with \nmt using the improved version of the Narrow Kernel Approximation (NKA \cite{1906.11765}) presented in \cite{Nicola_2020}, and using the measured power spectra in the estimation to avoid mis-modeling the signal or noise. We also included the contribution from super-sample covariance \cite{1302.6994}, computed with a halo model as described in \cite{1601.05779} and accounting for the variance of the linear density field $\sigma_B^2$ with the angular power spectra of the masks. Finally, we neglected the connected non-Gaussian covariance, which has a negligible effect for the scales and noise levels of these data \cite{1807.04266}. The method used to estimate both power spectra and covariances was able to reproduce the official DES measurements presented in \cite{2203.07128} at high precision when using the same band powers and pixelisation scheme.

  \subsection{Future cosmic shear data}\label{ssec:data.lsst}
    In Section~\ref{ssec:results.lsst} we will explore the ability of the methods presented here to avoid the difficulties of modelling the small-scale matter power spectrum in the context of upcoming Stage-IV surveys. To do so, we generate a synthetic data vector of power spectra mimicking the measurements that could be achieved for an imaging experiment such as the Vera Rubin Observatory's Legacy Survey of Space and Time (LSST). To do so, we follow the specifications for a 10-year survey outlined in the LSST Dark Energy Science Collaboration Science Requirements Document \cite{1809.01669}. As in \cite{2212.04291,2311.16812}, the full shear sample, with a total number density of $\bar{n}_{\rm gal}=26.5\,{\rm gal}/{\rm arcmin}^2$ was divided into five redshift bins containing approximately equal numbers of galaxies. The redshift distribution of each redshift bin was constructed by combining the model for the full redshift distribution presented in \cite{1305.0793} with a Gaussian photometric redshift error model with standard deviation $\sigma_z=0.05(1+z)$.

    The data vector contained all auto- and cross-correlations between these redshift bins in the range of multipoles $30\leq\ell\leq2000$. These were divided in 25 band powers as defined in \cite{2212.04291}. The covariance matrix was calculated including only the Gaussian or ``disconnected'' contributions, assuming a shape noise of $\sigma_e=0.28$ per ellipticity component. This neglects non-Gaussian contributions, which should be relatively subdominant given the large area covered by LSST. Furthermore, using optimistic uncertainties allows us to stress-test the ability of our method to avoid parameter biases due to a mis-modelling of the small-scale power spectrum, especially in the context of baryonic effects.

    To quantify the impact of these effects, we generate two versions of the synthetic data vector, using two different models for the matter power spectrum. In the first case we use a ``gravity-only'' matter power spectrum, which neglects the impact of baryonic effects. In the second case, we used a matter power spectrum including the impact of baryonic effects with \bemu as described in Section \ref{ssec:method.bar_effect}.

  \subsection{Likelihood}\label{ssec:data.like}
    \begin{table}[ht]
       \centering
       \begin{minipage}{0.55\textwidth} 
         \centering
         \begin{tabular}{| l  c |}
           \hline
           \multicolumn{2}{| c |}{\textbf{{Cosmology}}}\\
           $A_{s}10^9$ & $\mathcal{U}(0.50, 5.0)$ \\ 
           $\Om$ & $\mathcal{U}(0.10, 0.70)$ \\ 
           $\Omega_b$ & $\mathcal{U}(0.03, 0.07)$ \\ 
           $h$ & $\mathcal{U}(0.55, 0.91)$ \\ 
           $n_s$ & $\mathcal{U}(0.87, 1.07)$ \\ 
           $\sum m_{\nu}\,[{\rm eV}]$ & $\mathcal{U}(0.0559, 0.400)$\\
           \hline
           \multicolumn{2}{| c |}{\textbf{{\small IA model}}}\\
             $A_{IA}$& $\mathcal{U}(-5, 5)$\\ 
             $\eta_{IA}$&$\mathcal{U}(-5, 5)$\\
             \hline
               \multicolumn{2}{| c |}{\bemu}\\
             $\log_{10}(M_c)$& $\mathcal{U}(9, 15)$\\ 
             $\log_{10}(\eta)$&$\mathcal{U}(-0.7, 0.7)$\\
            $\log_{10}(\beta)$&$\mathcal{U}(-1.0, 0.7)$\\
            $\log_{10}(M_{\rm z0, cen})$&$\mathcal{U}(9, 13)$\\
            $\log_{10}(\theta_{\rm inn})$&$\mathcal{U}(-2, -0.53)$\\
            $\log_{10}(\theta_{\rm out})$&$\mathcal{U}(-0.48, 0)$\\
            $\log_{10}(M_{\rm inn})$&$\mathcal{U}(9, 13.5)$\\
            \hline
        \end{tabular}
    \end{minipage}%
    \hfill
    \begin{minipage}{0.41\textwidth} 
        \centering
        \begin{subtable}[t]{\linewidth}
            \centering
            \begin{tabular}{|p{1.5cm} p{3.5cm}|} 
            \hline
           \multicolumn{2}{| c |}{\textbf{{\small DES-Y3 shear calibration}}}\\
            $m_1$&$\mathcal{N}(-0.0063, 0.0091)$\\
            $m_2$&$\mathcal{N}(-0.0198, 0.0078)$\\
            $m_3$&$\mathcal{N}(-0.0241, 0.0076)$\\
            $m_4$&$\mathcal{N}(-0.0369, 0.0076)$\\
            \hline
               \multicolumn{2}{| c |}{\textbf{{\small DES-Y3 photo-z}}}\\
            $\Delta z_1$&$\mathcal{N}(0.0, 0.018)$\\
            $\Delta z_2$&$\mathcal{N}(0.0, 0.015)$\\
            $\Delta z_3$&$\mathcal{N}(0.0, 0.011)$\\
            $\Delta z_4$&$\mathcal{N}(0.0, 0.017)$\\           
            \hline
            \end{tabular}
        \end{subtable}
        \vspace{0.8cm} 
        \begin{subtable}[t]{\linewidth}
            \centering
            \begin{tabular}{|p{1.5cm} p{3.5cm}|} 
            \hline
           \multicolumn{2}{| c |}{\textbf{{\small LSST photo-z}}}\\
            $\Delta z_1$&$\mathcal{N}(0.0, 0.00115)$\\
            $\Delta z_2$&$\mathcal{N}(0.0, 0.0015)$\\
            $\Delta z_3$&$\mathcal{N}(0.0, 0.0017)$\\
            $\Delta z_4$&$\mathcal{N}(0.0, 0.002)$\\ 
            $\Delta z_5$&$\mathcal{N}(0.0, 0.0025)$\\
            \hline
            \end{tabular}
        \end{subtable}
    \end{minipage}
    \caption{Prior distributions for cosmological and nuisance parameters used for the different analyses carried out in this work. $\mathcal{U}(a,b)$ and $\mathcal{N}(a,b)$ represent a uniform and a Gaussian distribution respectively. Note that, for brevity, we have omitted the units of some parameters, namely $\log_{10}[M_c / (h^{-1} M_\odot)], \log_{10}[M_{z0\mathrm{,cen}} / (h^{-1} M_\odot)]$, and $\log_{10}[M_{\rm inn} / (h^{-1} M_\odot)]$.
    }
    \label{tab:priors}    
\end{table}
    To derive constraints on cosmological parameters from the real and synthetic data described in the previous sections, we make use of a Gaussian likelihood
    \begin{equation}\label{eq:likelihood}
      -2\log p({\bf d}|\vec{\theta}) \equiv \chi^2+K = \left({\bf t}(\vec{\theta})-{\bf d}\right)^T{\rm Cov}^{-1}\left({\bf t}(\vec{\theta})-{\bf d}\right)+K,
    \end{equation}
    with $K$ being an arbitrary constant; {\bf d}, the data vector containing all estimated power spectra; ${\bf t}(\vec{\theta})$, the theoretical prediction depending on the model parameters $\vec{\theta}$; and ${\rm Cov}$, the covariance matrix of ${\bf d}$.

    Our model depends on a number of free parameters. These include the cosmological parameters $\{A_s, n_s, \Om, \Omega_b, h, \sum m_{\nu}\}$, and intrinsic alignment parameters $\{A_{\rm IA}, \eta_{\rm IA}\}$, as well as nuisance parameter associated with shape measurement and photometric redshift uncertainties. Specifically, we marginalise over a multiplicative bias parameter $m_i$ and a redshift shift parameter\footnote{This was found to be a reasonable parametrisation to summarise photometric redshift uncertainties in \cite{2301.11978}, even in the context of Stage-IV surveys.} $\Delta z_i$ in each redshift bin. In some parts of the analysis presented in Section \ref{ssec:results.DES-Y3} we will also take into account the impact of baryonic effects in the matter power spectrum. In that case, we will use \bemu. As done in \cite{2403.13794}, when including baryonic effects we will marginalise over 7 additional parameters
    \begin{equation}
      \{\log_{10}(M_{\rm c}), \log_{10}(\eta), \log_{10}(\beta), \log_{10}(M_{\rm z_0, cen}), \log_{10}(\theta_{\rm inn}), \log_{10}(\theta_{\rm out}), \log_{10}(M_{\rm inn})\},
    \end{equation}
    introduced in Section~\ref{ssec:method.bar_effect} (here $M_{\rm c}, M_{\rm z_0,cen}$ and $M_{\rm inn}$ are expressed in units of $h^{-1} ~ \mathrm{M}_\odot$, which we do not include for brevity). The priors on all parameters used in this analysis are collected in Table~\ref{tab:priors}. The priors on $m_i$ and $\Delta z_i$ are the same as those used in \cite{Abbot_2021} for DES-Y3, and we use the requirements described in \cite{1809.01669} in the case of LSST.

    With the scale cuts described in Sections \ref{ssec:data.descls} and \ref{ssec:data.lsst}, the total number of data points is $N_d=240$ for DES-Y3 and $N_d = 270$ for LSST. To sample the parameter space, we make use of the {\tt Cobaya}\footnote{\url{https://cobaya.readthedocs.io/en/latest}}  code \cite{Torrado_2020}, employing its implementation of the Metropolis-Hasting Markov Chain Monte Carlo (MCMC, \cite{metropolismc, hastingsmc}) algorithm. We considered parameter chains to have converged when the Gelman-Rubin metric \cite{gelmanrubin} $R$ reached $R-1\leq 0.03$. All theoretical predictions were obtained using the {\tt Core Cosmology Library}\footnote{\url{https://ccl.readthedocs.io/en/latest/}} \cite{Chisari_2018}, with the non-linear matter power spectrum computed using \bemu \cite{Arico_2021,AnguloEtal2021}.

\section{Results}\label{sec:res}
  \subsection{DES-Y3 analysis}\label{ssec:results.DES-Y3}
    \subsubsection{Power spectra}\label{sssec:results.DES-Y3.cls}
      \begin{figure}[t]
        \centering        
        \hspace{-0.6cm}\includegraphics[width = \textwidth]{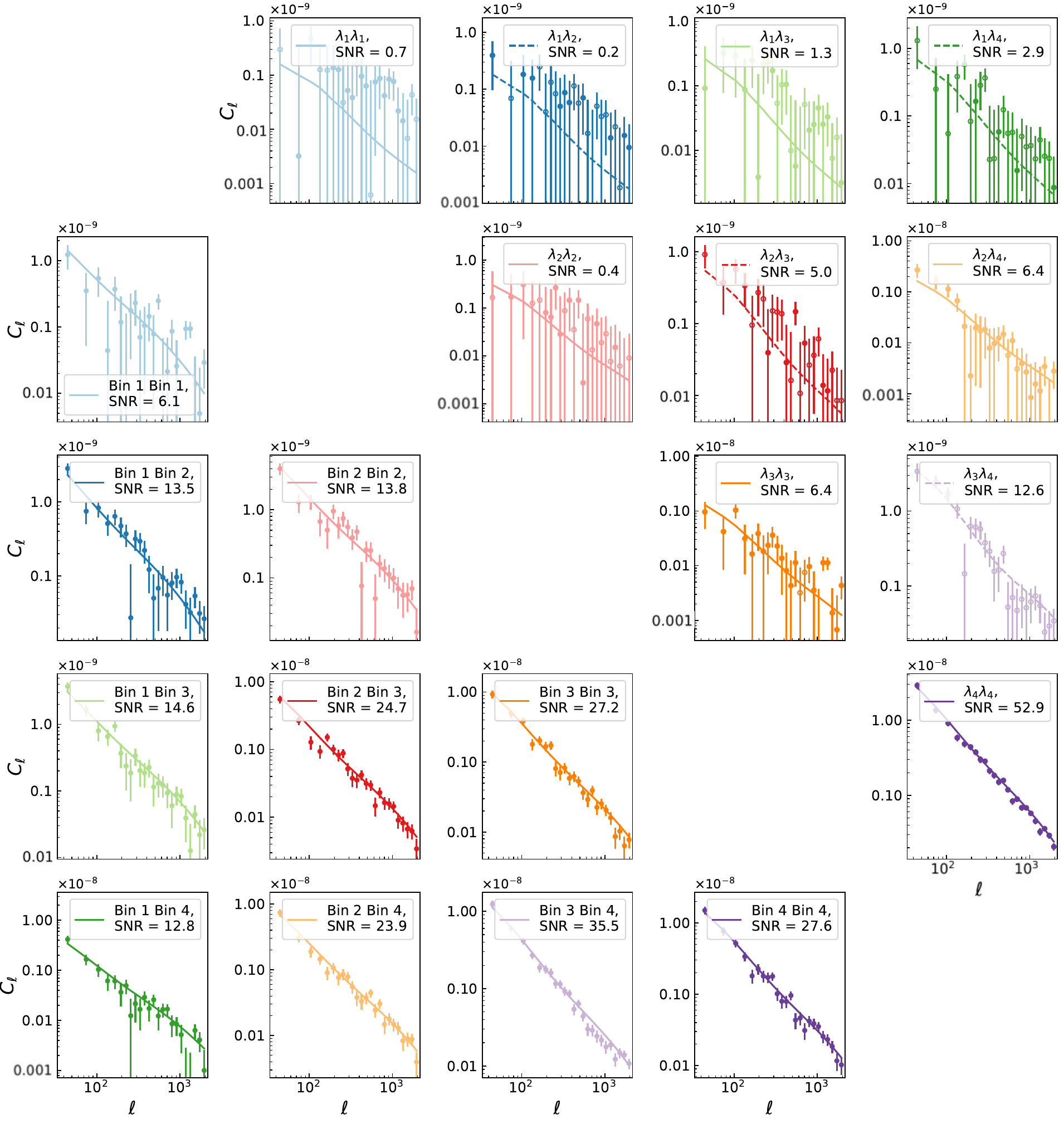}
        \caption{{\sl Lower triangle}: In coloured dots, the estimated power spectra for the DES-Y3 bins. {\sl Upper triangle}: In coloured dots, the rotated data vector obtained when applying the \dnull for $\chi_{L} = 1000~{\rm Mpc}$ to the input power spectra. In both cases, lines represent the theoretical predictions obtained with the best fitting parameters of all data and marginalizing over baryonic effects (but without including IAs). Note that empty dots and dashed lines refer to negative value of the $C_{\ell}s$.
        In each plot, we list the expected signal-to-noise ratio (SNR) from each bin computed with Equation~\ref{eq:snr_est}.}\label{fig:cls_desy_rot_notelldep}
      \end{figure}

      The power spectra measured from the DES-Y3 data are shown in the lower-left panels of Figure~\ref{fig:cls_desy_rot_notelldep}. The upper right panels of the same figure show the auto- and cross-correlations of the nulling eigenmodes, defined via \dnull with $\chi_L=1000\,{\rm Mpc}$, which corresponds roughly to the median radial distance of the first redshift bin. Solid lines in the panels of the lower triangle represent the theoretical predictions obtained for the DES-Y3 bins angular power spectra, derived using the best-fit parameters found using all the data. Similarly, the solid and dashed lines in the upper triangle correspond to the rotated version of the theoretical predictions obtained after applying the \dnull transformation. From these measurements we can see clearly that the higher eigenmodes (e.g. $\lambda_4$ and $\lambda_3$), which are the most sensitive to small-scale structures at low redshifts, also have the highest signal-to-noise ratio. This is as expected, given the relative sizes of the effective radial kernels shown in the top right panel of Figure~\ref{fig:qi_qw_desy}. In each of the panels, we list the signal-to-noise ratio (SNR) for each individual power spectrum. This was calculated by fitting the measurements to a theoretical prediction of the form  ${\bf t}=b\,{\bf t}_*$, where $b$ is a free amplitude parameter and ${\bf t}_*$ is a free template (given by the best-fit theory prediction above). The SNR is then given as
      \begin{equation}\label{eq:snr_est}
        {\rm SNR}\equiv\frac{\hat{b}}{\sigma_b}=\frac{{\bf t}^T_*{\rm Cov}^{-1}{\bf d}}{\sqrt{{\bf t}_*^T{\rm Cov}^{-1}{\bf t}_*}},
      \end{equation}
      where $\hat{b}$  is  the best-fit value of $b$ and $\sigma_b$ its standard deviation\footnote{This method is more appropriate than the standard approximation ${\rm SNR}\simeq\sqrt{{\bf d}^T{\rm Cov}^{-1}{\bf d}}$ for noise-dominated datasets, as is the case for the lowest eigenmodes.}.
      \begin{figure}[h]
        \centering
        \includegraphics[width = 0.9\linewidth]{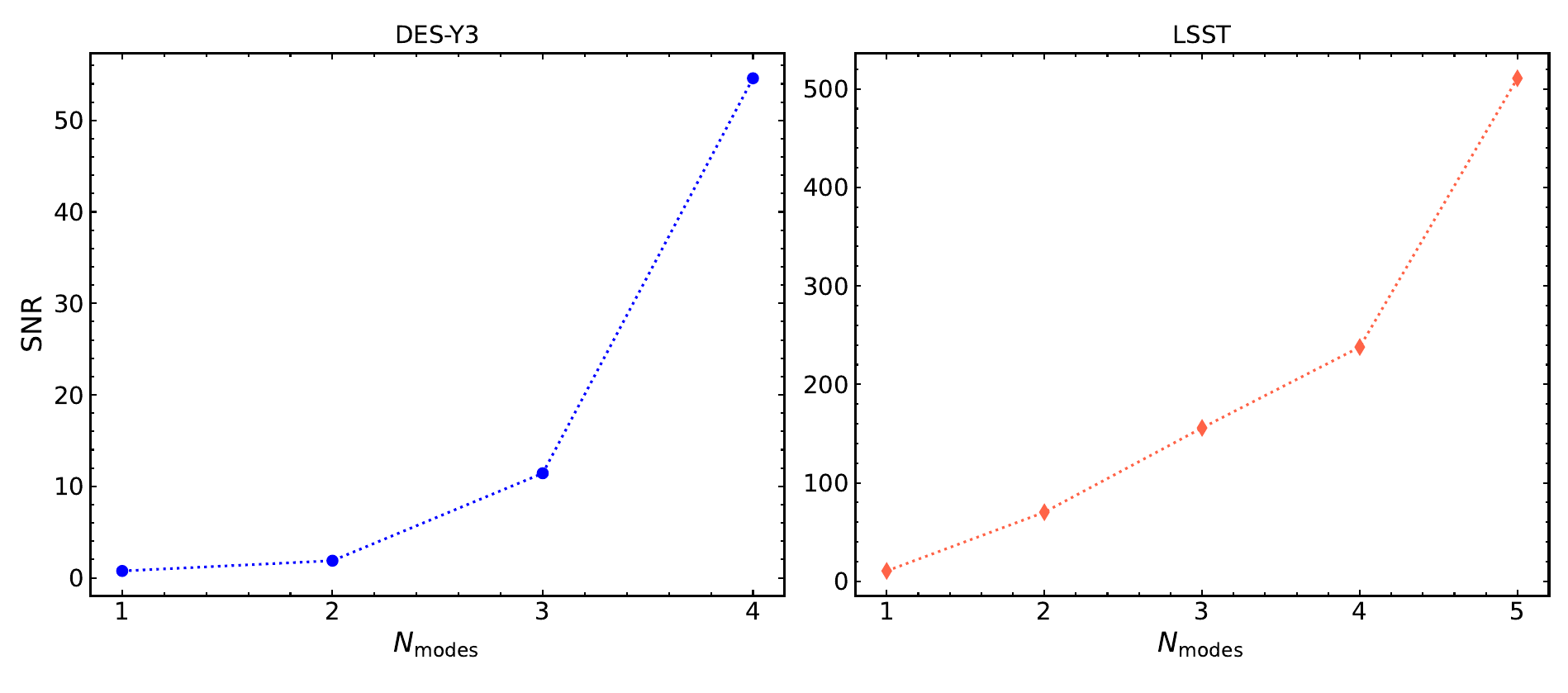}
        \caption{Total signal-to-noise ratio SNR as a function of the number of modes considered in the analysis for DES-Y3 ({\sl left panel}) and LSST ({\sl right panel}).}\label{fig:snr_nmodes}
      \end{figure}
      To further quantify the amount of signal contained in each eigenmode, we calculate the cumulative SNR for different sets of eigenmode combinations. This is shown in the left panel of Figure \ref{fig:snr_nmodes} as a function of the number of eigenmodes included in the data, starting with the least contaminated eigenmode (e.g. for $N_{\rm modes}=3$ the data vector contains all auto- and cross-correlations between the first three eigemodes $\{\lambda_1,\lambda_2,\lambda_3\}$). We can see that the total SNR is dominated by the most contaminated mode, and excluding it reduces the SNR by $\sim70\%$.

      The results for the other \dnull and \knull combinations explored in later sections are similar to those presented here for \dnull with $\chi_L=1000\,{\rm Mpc}$ and we omit them for brevity. The main quantitative differences between them are small and better discussed in terms of their parameter constraints, which we proceed to describe.

    \subsubsection{Constraints on cosmological parameters}\label{sssec:results.DES-Y3.cosmo}
      \begin{figure}[h!]
        \centering
        \includegraphics[width=\textwidth]{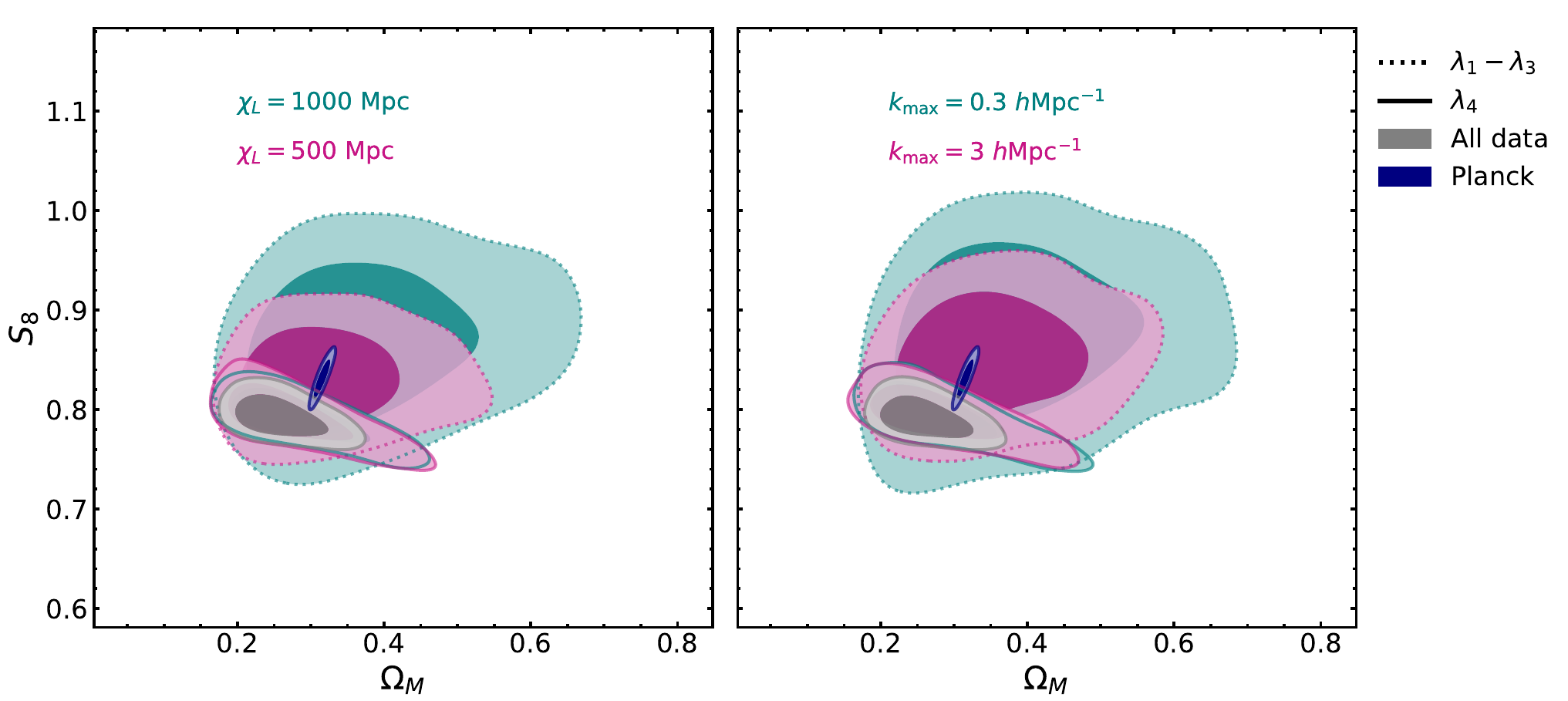}
        \caption{Constraints in the $S_8$ and $\Om$ plane (68\% and 95\% C.L.) for the \dnull case ({\sl left panel}) and \knull case ({\sl right panel}) from DES-Y3. In both cases, navy contours show the results obtained from \planck \cite{1807.06209}. Coloured contours show results for the  four different number of eigenmodes and values of $\chi_{L}$ and $k_{\rm max}$, respectively. Contours having the same colour are shown with two linestyles: dotted lines are results from the runs with the first three eigenmodes ($\lambda_1- \lambda_3$) while solid lines correspond to using the highest mode alone ($\lambda_4$). The different values of $\chi_L$ and $k_{\rm max}$ are listed in each panel. Note that, for this test, we marginalised over all BCM parameters while we did not include contribution from IAs.}
        \label{fig:desy_params_chimin}
      \end{figure}
      We have seen that the total cosmic shear signal in the DES-Y3 data is dominated by the most contaminated mode, $\lambda_4$. However, Figure \ref{fig:snr_nmodes} shows that the shear signal from the remaining 3 modes is still detected above $10\sigma$. It is therefore interesting to consider the impact on the current cosmological constraints from cosmic shear of removing this mode. To do this, we perform several MCMC runs, each considering a different combination of modes, and applying different modelling choices. In all cases we use the same scale cuts and, by default, we ignore intrinsic alignments (the impact of IA is discussed later), but take into account baryonic effects, marginalising over all the BCM parameters. When applying the \dnull transformation, we considered distance cuts at $\chi_L = [500, 1000, 1500] ~{\rm Mpc}$ to quantify the sensitivity of our results to the width of the redshift distribution in the first bin. In turn, for \knull we considered scale cuts $k_{\rm max} = [0.3, 1, 3] ~ h\,{\rm Mpc}^{-1}$, which cover the range of scales where baryonic effects are most relevant \cite{2410.18191}.
      
      The left panel of Figure~\ref{fig:desy_params_chimin} presents an example of the cosmological constraints on $S_8$ and $\Om$ (the main parameters constrained by cosmic shear data) obtained using the \dnull approach. Results are shown for $\chi_{L}=500$ and $1000~{\rm Mpc}$ (we present the results for $1500~{\rm Mpc}$ later on). We consider three different combinations of eigenmodes: the most-contaminated mode alone, $\lambda_4$, the first three least contaminated modes ($\lambda_1 - \lambda_3$), and the combination of all modes (which, by construction, is equivalent to using the full data vector of cosmic shear power spectra between all pairs of redshift bins). The figure also shows the CMB constraints on these parameters from \planck \cite{1807.06209}. The constraints found with the full data vector agree with those presented in \cite{2403.13794} when using the same data. In this case, the data favours a value of $S_8$ that is lower than the CMB value at the $2.6\sigma$ level. Repeating the analysis using only the most contaminated mode leads to $\sim 60\%$ broader constraints on $\Om$, and constraints on $S_8$ that are almost equivalent to the full-data case. This again shows that this mode dominates the constraining power of the data. It is interesting that the addition of the $(\lambda_1,\lambda_2,\lambda_3)$ modes mostly improves the constraints on $\Om$, with a significantly milder impact on $S_8$. This may be explained by the fact that, since $\lambda_4$ dominates the overall SNR of the data, it is able to maximise the constraining power on the most important amplitude-like parameter ($S_8$). The additional modes are then necessary to extract information about the redshift evolution of this amplitude (i.e. the growth of structure), which in turn is governed by $\Om$. Removing the most contaminated mode (i.e. keeping $(\lambda_1,\lambda_2,\lambda_3$)) leads to significantly broader constraints on both parameters, as expected given the significant loss of signal that dropping $\lambda_4$ entails (see Figure \ref{fig:snr_nmodes}). As could be expected from Figure \ref{fig:qi_qw_desy}, using a more conservative $\chi_{L}$ (i.e. $\chi_{L}=1000\,{\rm Mpc}$ instead of $\chi_L=500\,{\rm Mpc}$) transfers information from the least contaminated modes to the most contaminated one, and thus the constraints from $\lambda_4$ alone become tighter, while those from $\lambda_1 - \lambda_3$ become broader.

    \begin{table}[t]
    \renewcommand{\arraystretch}{1.5}
    \centering
    \begin{tabular}{|c|c|c c||c|c c|}
    \hline
    \multirow{3}{*}{Modes} & \multicolumn{3}{c||}{\dnull} & \multicolumn{3}{c|}{\knull} \\
    \hline
    & $\chi_L$ & \multirow{2}{*}{$S_8$} & \multirow{2}{*}{$\Om$} & $k_{\rm max}$ & \multirow{2}{*}{$S_8$} & \multirow{2}{*}{$\Om$} \\
    & ${\small [{\rm Mpc}]}$ & & & ${\small [h{\rm Mpc}^{-1}]}$ & & \\
    \hline
    $\lambda_1$-$\lambda_3$ & $500$ & $0.850 \pm 0.043$ & $0.355 \pm 0.091$ & $0.3$ & $0.864 \pm 0.055$ & $0.369 \pm 0.105$ \\
     & $1000$ & $0.874 \pm 0.060$ & $0.383 \pm 0.107$ & $1$ & $0.835 \pm 0.043$ & $0.333 \pm 0.079$ \\
     & $1500$ & $0.844 \pm 0.074$ & $0.383 \pm 0.110$ & $3 $ & $0.830 \pm 0.035$ & $0.321 \pm 0.074$ \\
     \hline
    $\lambda_4$ & $500 $ & $0.793 \pm 0.019$ & $0.280 \pm 0.056$ & $0.3$ & $0.792 \pm 0.019$ & $0.285 \pm 0.062$ \\
     & $1000$ & $0.796 \pm 0.018$ & $0.272 \pm 0.055$ & $1$ & $0.799 \pm 0.017$ & $0.254 \pm 0.046$ \\
     & $1500$ & $0.794 \pm 0.017$ & $0.285 \pm 0.064$ & $3$ & $0.794 \pm 0.019$ & $0.269 \pm 0.052$ \\
     \hline
        \end{tabular}
        \caption{Constraints on $S_8$ and $\Om$ from different choices of nulling scheme and small-scale/distance removal criteria when neglecting intrinsic alignments and baryonic effects. The constraints using all modes are $S_8=0.790\pm0.010$ and $\Om = 0.258^{+0.028}_{-0.052}$. Note that $\lambda_1 -\lambda_3$ refers to the combination of modes given by: $\lambda_1 +\lambda_2+\lambda_3$.}
        \label{tab:s8_des}
      \end{table}

      \begin{figure}[h!]
        \centering
        \includegraphics[width=0.9\linewidth]{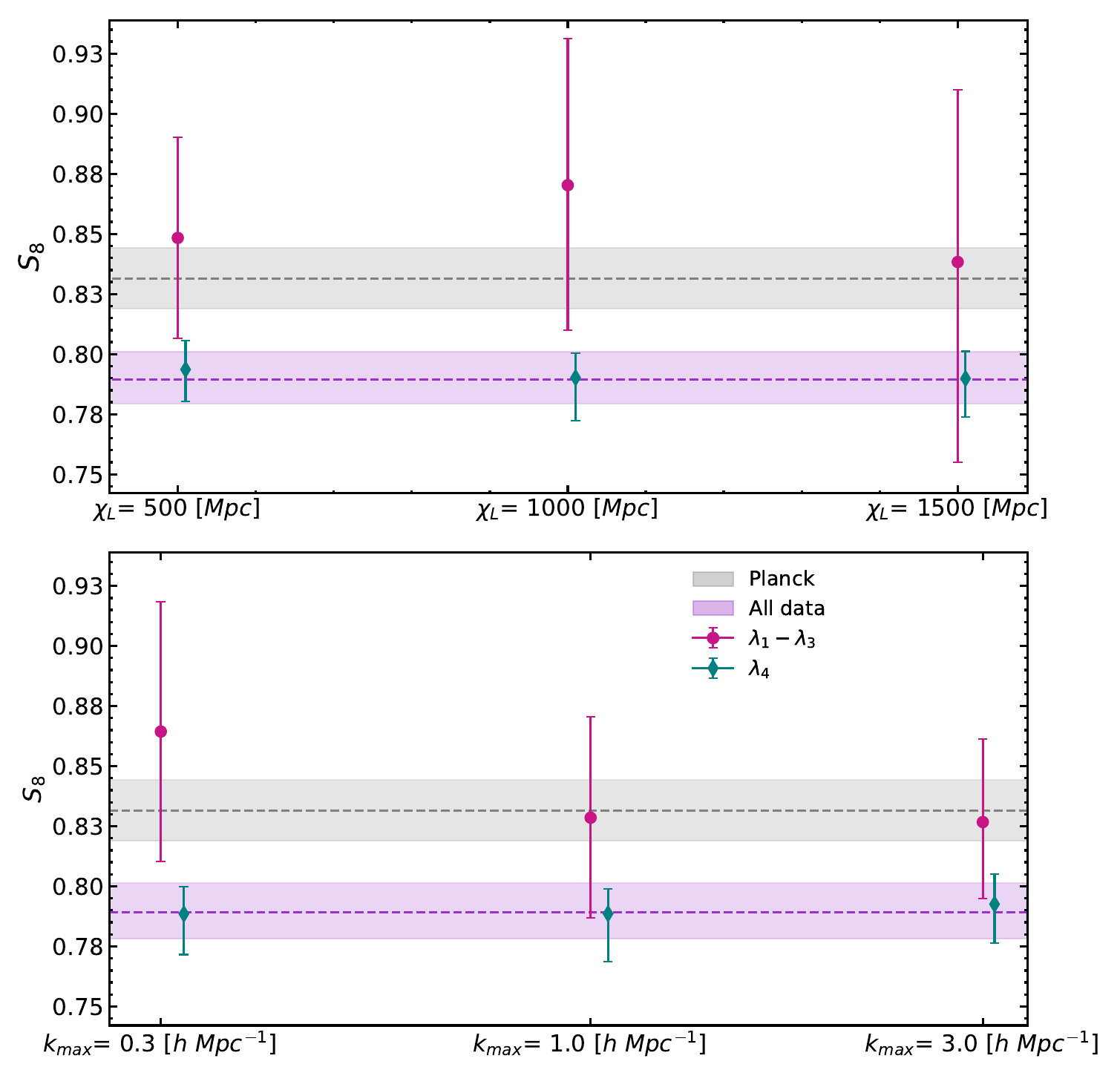}
        \caption{Estimated values of $S_8$ as a function of $\chi_{L}$ ({\sl top panel}) and $k_{\rm max}$ ({\sl bottom panel}) when neglecting intrinsic alignments and baryonic effects. The gray band represents the 1$\sigma$ constraints from \planck \cite{1807.06209}. The purple shadowed area represents the  $1\sigma$ constraints obtained when considering the full data vector (i.e. including all eigenmodes, $\lambda_1 - \lambda_4$), while coloured markers represent constraints on $S_8$ for different combinations of modes. The pink points show the results found using the first three least contaminated modes, while the green markers show the results obtained from the most contaminated mode. A horizontal shift is applied to the green markers to ease the visualisation. }
        \label{fig:s8_DES-Y3}
      \end{figure}

      Interestingly, besides the increase in uncertainties, restricting ourselves to $(\lambda_1  - \lambda_3)$ also leads to an upwards shift in the preferred value of $S_8$. To understand the significance of this shift, and the potential evidence of a systematic mis-modelling of the data on small scales it could entail, we repeat this analysis for different values of $\chi_L$ in the \dnull approach, and different values of $k_{\rm max}$ for the \knull case. To simplify the interpretation of the result in terms of small-scale modelling, we neglect baryonic effects and IA in this analysis. The result of this study is shown in Figure~\ref{fig:s8_DES-Y3} and Table \ref{tab:s8_des}. The horizontal bands in the figure show the constraints on this parameter found by \planck ($S_8=0.832\pm 0.013$), and those obtained from our DES-Y3 data vector when including all modes ($S_8= 0.790 \pm  0.010$). 
      The constraints obtained using only the most contaminated mode are very similar to the all-modes case, displaying a similar downwards shift with respect to \planck, with only $\sim 16\%$ larger error bars. The constraints obtained discarding the most contaminated mode, in contrast, are shifted upwards by $1\sigma$, displaying a closer agreement with \planck. In the case of \dnull, we see a clear increase in uncertainties as we move to more conservative distance cuts, particularly for $\chi_{L}=1500\,{\rm Mpc}$, which is beyond the distance to the median redshift of the first DES-Y3 bin. The uncertainties for $\chi_{L}=500\,{\rm Mpc}$ are $\sim 3.4$ times larger than those obtained using the full data vector, and they grow by an additional extra factor of $\sim 1.8$, resulting in $~6.2$, for $\chi_{L}=1500\,{\rm Mpc}$. In the case of \knull we observe a similar shift, with a factor $\sim 5.5$ increase in uncertainties in the most conservative case ($k_{\rm max}=0.3\,h~{\rm Mpc}^{-1}$), which shrinks to $\sim 3.5$ in the least conservative case ($k_{\rm max}=3\,h~{\rm Mpc}^{-1}$).

      A priori, it is not obvious whether the shift we observe in $S_8$ when dropping the most contaminated mode is a statistically significant indication of a modelling systematic on small scales. Although the shift is only $\sim1$-$2\sigma$ away from the value of $S_8$ obtained when using all the data, both measurements have data in common, and thus their statistical uncertainties are correlated to some extent. To better quantify this, we carry out the following frequentist test: we generate 100 simulations of the data vector containing all shear auto- and cross-correlations. These are generated from a multi-variate Gaussian distribution with a mean given by the best-fit theory prediction found from the full dataset, and the covariance matrix of our data. Then, for each of these simulations, we find the best-fit model parameters by maximising the likelihood of Equation~\ref
      {eq:likelihood} using the full data vector, and using only the three least-contaminated eigenmodes. We do so only for the \knull case with $k_{\rm max}=0.3\,h~{\rm Mpc}^{-1}$, for simplicity. We then study the distribution of the difference in the value of $S_8$ found in these two cases $\Delta S_8\equiv S_8^{\lambda_1\textrm{-}\lambda_3}-S_8^{\lambda_1\textrm{-}\lambda_4}$ (where $S_8^{\lambda_i\textrm{-}\lambda_j}$ indicates that we considered the modes from $\lambda_i$ to $\lambda_j$). We find that 17\% of all simulations recovered upwards shifts in $S_8$ between these two cases that are larger the value we find in the data, and that 30\% of all simulations recover larger absolute differences between the two best-fit values of $S_8$, regardless of sign. Our initial expectation is therefore correct: the shift observed in $S_8$ when removing the most contaminated mode is only significant at the 1-2$\sigma$ level, and therefore cannot be interpreted as a clear indication of systematic mis-modelling of the signal on small scales.

      All the results presented thus far were obtained assuming no IA contribution in the model. To investigate the effect of IA on the cosmological constraints, we run MCMC chains for the \dnull case with $\chi_L = 1000\,{\rm Mpc}$ including the two IA parameters $A_{\rm IA}$ and $\eta_{\rm IA}$ adopting the priors listed in Table~\ref{tab:priors}. The general conclusions we draw from the analysis summarised in Figure~\ref{fig:s8_DES-Y3} remain qualitative unchanged after including IA. As a result of introducing these two additional free parameters, the constraining power is generally reduced, leading to a degradation in the errorbars of the cosmological parameters. In particular, comparing with the results found without IA, the inclusion of IA results in an increase $S_8$ estimate up to 27\% and 32\% for $\lambda_1-\lambda_4$ and $\lambda_1-\lambda_3$, respectively. When removing the highest eigenmode, we still observe an upward shift in the $S_8$ constraint and a sigificant increase in the error bars with respect to the $\lambda_1-\lambda_4$ case. In contrast with the results found without IA, the error bars obtained using only $\lambda_4$ increase by a factor $\sim 3.7$ with respect to the full dataset. This is because, in the absence of tomographic information, it is not possible to disentangle the IA and the lensing contributions using their different redshift structure (IA are local in redshift, whereas lensing is cumulative), making $A_{\rm IA}$ and $S_8$ effectively degenerate. We discuss this in more detail in the next section.

\subsection{Future Stage-IV shear surveys}\label{ssec:results.lsst}
  \begin{figure}[t]
    \centering
    \includegraphics[width = \textwidth]{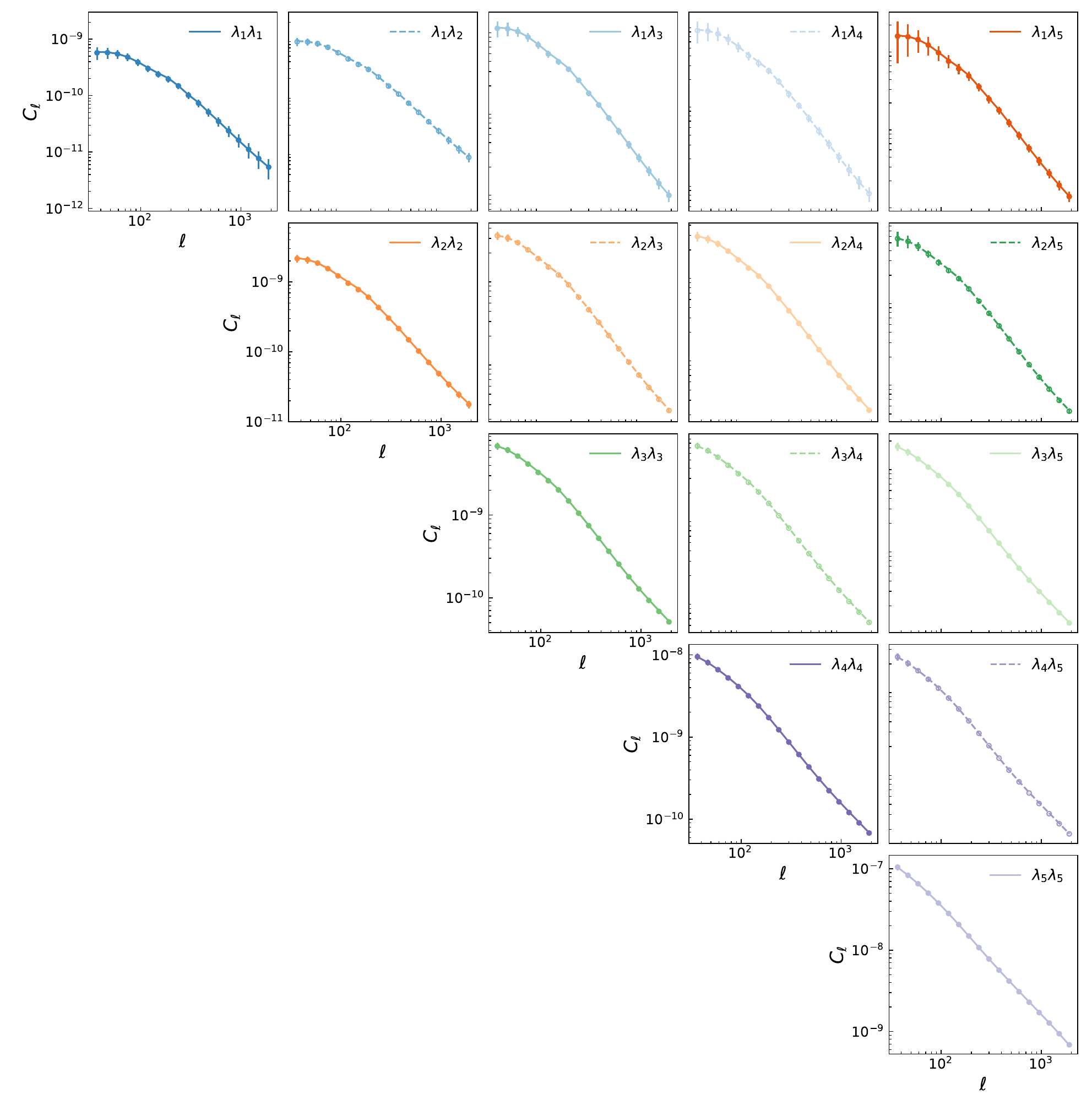}
    \caption{Same as Figure~\ref{fig:cls_desy_rot_notelldep}, but for the five LSST bins in the \texttt{baryons} case. The figure displays the measured power spectra (full and empty dots) and theoretical predictions (solid and dashed lines) after applying the \dnull transformation with $\chi_{L} = 1000 ~{\rm Mpc}$. The input theory predictions (not shown in this plot) are computed with the best fitting parameters to all data, without marginalizing over IA nor BCM parameters.}
    \label{fig:cls_lsst_rot_notelldep}
  \end{figure}
  We performed a similar analysis to the one described in Section~\ref{sssec:results.DES-Y3.cosmo}, using simulated LSST Y10 data. For this, we use the five redshift bins, number density, and scale cuts described in Section \ref{ssec:data.lsst}. To quantify the impact of baryonic effect mis-modelling (as discussed in Section~\ref{ssec:method.bar_effect}), we repeat this analysis for two different data vectors of cosmic shear power spectra, one including the impact of baryonic effects for the fiducial BCM model described in Section \ref{ssec:method.bar_effect}, and one without baryonic effects. 

  To begin with, we study the contribution to the overall weak lensing signal from different nulling eigenmodes. Figure~\ref{fig:cls_lsst_rot_notelldep} presents the rotated angular power spectrum after applying the \dnull transformation for $\chi_L = 1000~{\rm Mpc}$. Unlike in the case of DES-Y3, we can see that, although the mode that is most sensitive to non-linear scales $(\lambda_5)$ still achieves the highest signal-to-noise ratio, all modes, including the least contaminated one ($\lambda_1$), are detected at high significance. The cumulative SNR for these data is shown in the right hand side of Figure~\ref{fig:snr_nmodes}. The first (cleanest) eigenmode is detected at $\sim10\sigma$, with the total SNR raising to $\sim500$ when including all modes. This SNR drops by a factor $\sim2$ after removing the most contaminated mode.

  Using this synthetic data vector, we derive constraints on cosmological parameters from different subsets of eigenmodes, using the same procedure outlined in Section \ref{ssec:results.DES-Y3} for the DES-Y3 data. For simplicity, we  consider only the \knull transformation with scale cuts $k_{\rm max} = [0.3, 1, 3] ~ h ~ {\rm Mpc}^{-1}$, and explore two combinations of modes: ($\lambda_1-\lambda_5$),  corresponding to the full data vector, and ($\lambda_1-\lambda_4$), where we remove the most contaminated mode. Unlike the DES-Y3 analysis, we always excluded from the theory predictions the presence of the baryons when computing the matter power spectrum. This resembles more closely a scenario in which we wish to be completely agnostic about baryonic physics, and aim to find the combinations of our data that allow us to forego modelling baryonic effects altogether. We repeat this analysis on the synthetic data vector containing baryonic effects, and on the one without them. The lack of baryons in the theory predictions will result in a bias on the inferred cosmological parameters in the former case. 
  
  \begin{figure}
    \centering
    \includegraphics[width=0.9\linewidth]{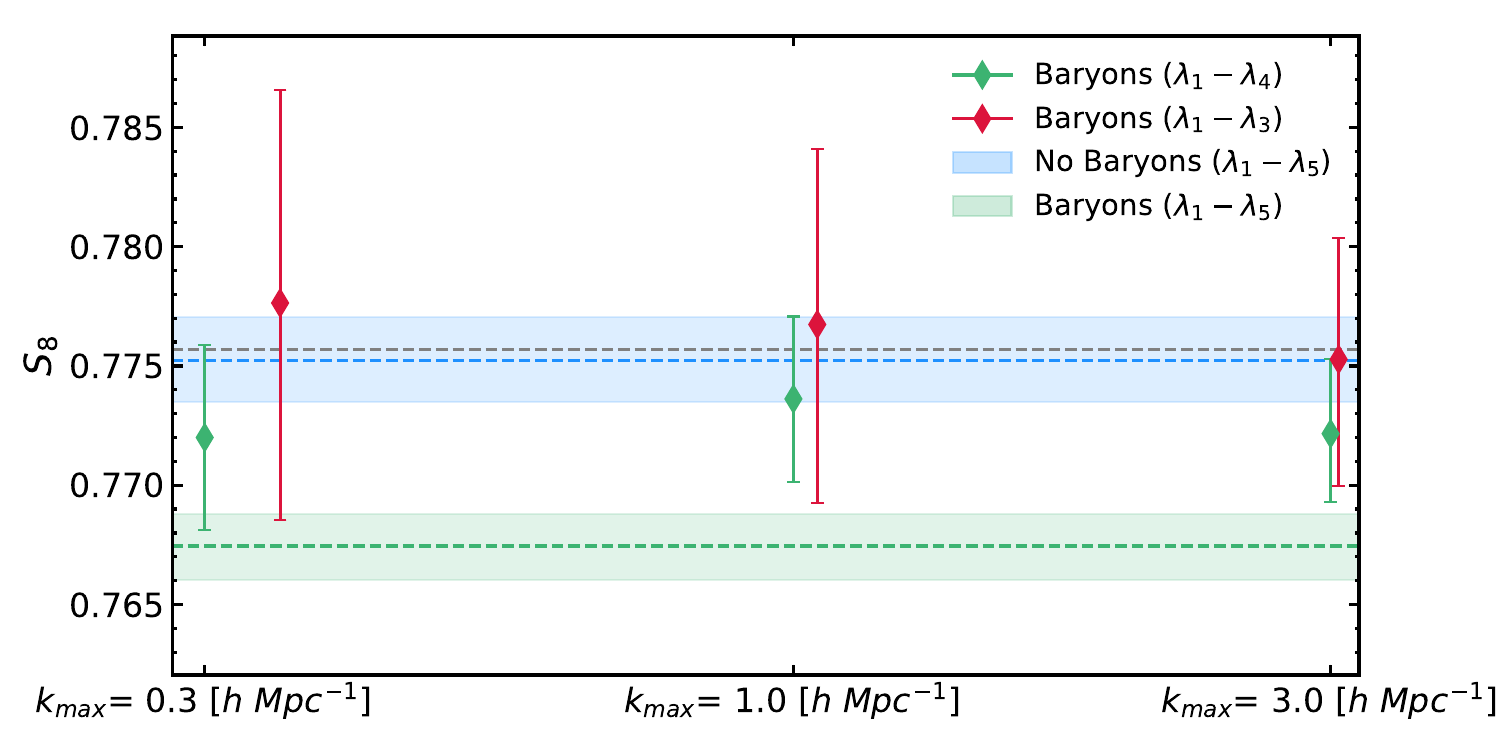}
    \caption{$S_8$ $1\sigma$-constraints when analysing a data vector that does not contain baryonic effects (blue band) or contains them (as described in Section \ref{ssec:method.bar_effect}, green band), with a model that neglects them. The green and red points with error bars show the constraints obtained after discarding the most contaminated modes (in the case with baryons in the dataset), using the \knull framework, as a function of $k_{\rm max}$. The green and blue dashed lines show the mean value of $S_8$ and the gray line, its input value. In these two cases we are able to reduce the $\sim5\sigma$ bias caused by ignoring baryonic effects to $\sim1\sigma$ or less. However, this is at the cost of a significant increase in the final statistical uncertainties. Note that, as in all tests of this Section, we did not include baryonic effects in the model and, for this case, no intrinsic alignments were considered.}
    \label{fig:s8_kmax_lsst}
  \end{figure}
  As in the DES-Y3 analysis, our primary focus is on the constraints on $S_8$, the parameter that is measured with the highest precision. These constraints are shown in Figure~\ref{fig:s8_kmax_lsst} as a function of the considered $k_{\rm max}$ values. The light blue horizontal band shows the $1\sigma$ region found when analysing the gravity-only data vector. By construction, in this case we are able to recover the value of $S_8$ from the input cosmology (shown with the dashed grey line). The green horizontal band, in turn, shows the constraints obtained when analysing the full data vector containing baryonic effects. The resulting value of $S_8$ is $\sim 5.6\sigma$ lower than the input value, with uncertainties that are similar to those obtained in the unbiased case. Excluding the most contaminated mode, $\lambda_5$, we observe a behaviour similar to that found in Section~\ref{sssec:results.DES-Y3.cosmo}: the preferred value of $S_8$ (green points with error bars in the figure) experiences an upwards shift towards the true value of $S_8$, accompanied by an increase in the statistical uncertainties by a factor of $2.8$ for $k_{\rm max} = 0.3~{\rm h Mpc^{-1}}$, due to the loss of sensitivity caused by dropping $\lambda_5$. Removing the single most contaminated mode thus reduces the bias in $S_8$ from $5.6\sigma$ to $\lesssim1\sigma$, albeit at the cost of a $\sim3$-fold increase in the statistical uncertainties. Furthermore, we considered the case where we remove the two most contaminated modes and constrain $S_8$ from the combination of modes $\lambda_1 - \lambda_3$. This is reasonable, considering that the signal from these remaining modes is still detected with a high significance for this data vector ($\sim 160\sigma$, see Figure~\ref{fig:snr_nmodes}). Results for this test are shown with red markers in Figure~\ref{fig:s8_kmax_lsst}. The residual $\sim1\sigma$ bias in $S_8$ present in the 4-mode case now disappears completely, albeit at the cost of an additional increase of the statistical uncertainties by a factor of 2.1.

\begin{figure}
    \centering
    \includegraphics[width = \linewidth]{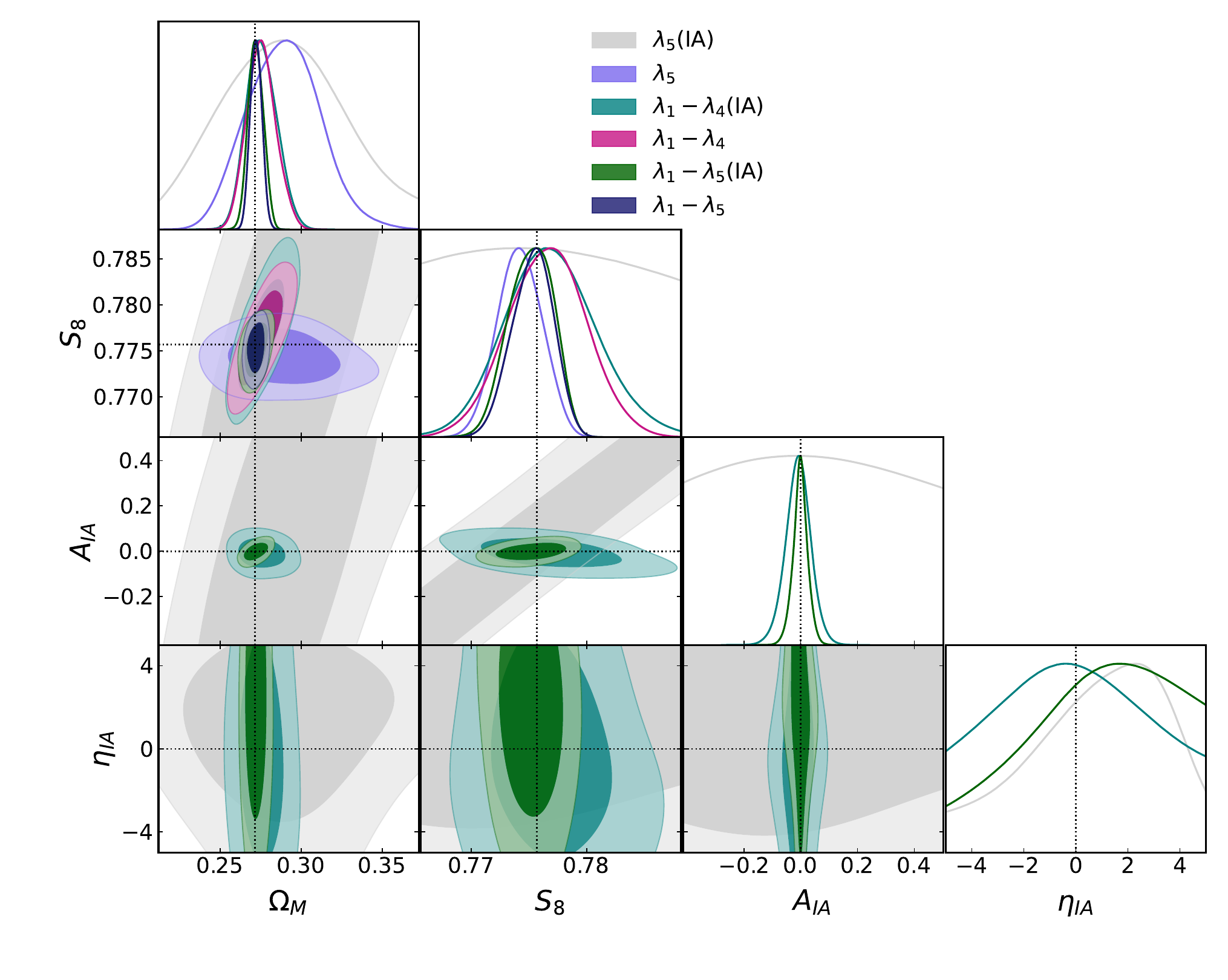}
    \caption{Constraints on cosmological parameters from the LSST power spectra in the {\tt no baryons} case when applying the \knull transformation with $k_{\rm max} = 1~{\rm h Mpc^{-1}}$. Coloured contours show outcomes for different combinations of eigenmodes when marginalizing or not over the contribution of IAs. Note that for these runs, we did not include BCM parameters. Dotted lines mark the input cosmology and IA parameters for all of the runs.}
    \label{fig:lsst_ia}
  \end{figure}

  As discussed in Section~\ref{sssec:results.DES-Y3.cosmo}, the presence of intrinsic alignments in the theory model has an impact on the cosmological constraints that depends on the set of modes included in the analysis. To explore this further, and to better understand the parameter degeneracies of different data subsets, we study the constraints in the $\{S_8,\Om,A_{\rm IA},\eta_{\rm IA}\}$ parameter space obtained from different mode combinations, with and without intrinsic alignments. The results are shown in Figure~\ref{fig:lsst_ia}.
  When considering the full data vector, the inclusion of IA leads to a relatively small increase in the final statistical uncertainties on $S_8$ and $\Om$ (dark blue and green contours in the figure). This is not entirely surprising, given the relatively simple IA model used in our analysis and the high sensitivity of the simulated data. Future Stage-IV surveys will likely require more sophisticated IA parametrisations to ensure robustness against this contamination \cite{1708.09247,2307.13754,2311.16812}. Ignoring the impact of IA, it is still possible to constrain cosmological parameters using either the most contaminated mode, $\lambda_5$ (light purple contours), or the remaining modes (pink contours). Interestingly, we find a more pronounce version of the result found for DES-Y3: the most-contaminated mode is able to constrain $S_8$ at high precision, but it is significantly less sensitive to $\Om$. In turn, the remaining modes are able to constrain $\Om$ well, but are $\sim3$ times less sensitive to $S_8$ (as we found in the DES-Y3 case). The constraints found in these two cases are largely orthogonal to one another, and their combination results in the highly precise final constraints found for the full data vector. As we argued in Section \ref{sssec:results.DES-Y3.cosmo}, $\lambda_5$ is most sensitive to amplitude parameters, such as $S_8$, while the combination of all other modes is more sensitive to geometry and structure growth, and hence to $\Om$.
  
  These results change significantly in the presence of IA. As shown by the gray contours, $\lambda_5$ alone loses all constraining power on either $S_8$ or $\Om$. This is because, in the absence of tomography, both amplitude parameters ($A_{\rm IA}$ and $S_8$) are almost completely degenerate, as shown in the $(A_{\rm IA},\,S_8)$ sub-panel of the figure. In turn, the constraints found from the combination of all remaining modes (teal contours) are only mildly degraded (parameter uncertainties grow by $\sim 18\%$) by the presence of IA. This is because the availability of tomographic information in this mode combination allows us to disentangle the IA and lensing contributions, and to measure both $A_{\rm IA}$ and $\eta_{\rm IA}$ with reasonable precision. Combining this with $\lambda_5$ breaks the $A_{\rm IA}-S_8$ degeneracy mentioned above, allowing us to measure all parameters with errors that are only $\sim7\%$ larger than those found without IA.

\section{Conclusions}\label{sec:concl}
  Weak lensing analyses provide a strong framework to probe the clustering of matter with statistically powerful samples and test our current cosmological models. However, the difficulty of modelling small non-linear scales (i.e. because of the effect of non-linear gravitational evolution, astrophysical processes, or intrinsic alignments) can introduce large uncertainties that hinder their robustness. This is even more important for near-future Stage-IV surveys. In particular, while adopting ad-hoc scale cuts can help  reducing these uncertainties, the contribution of low-redshift structure (probing small, non-linear scales) is more difficult to mask out of our analyses \cite{2410.18191}. One class of methods developed for this purpose are the so called ``nulling'' techniques -- modifications to weak lensing kernels that make it possible to define angular scale cuts closely reflecting physical scale cuts.

  In this work we generalise these approaches by defining a linear combination of the tomographic bins of a survey that effectively suppresses the contribution of low-redshift structures. We identify two ways of achieving this objective: the first one, suppressing the contribution of structures below a given comoving distance, which we call \dnull (Equation \ref{eq:R}); the second one, which we call \knull, suppressing the contributions coming from scales smaller than a given wavenumber $k_{\rm max}$ (Equation \ref{eq:R-knull}). This second approach allows us to more directly connect the scales that are excluded from the analysis to the scales on which current models (for example of the full-physics matter power spectrum) break down.

  We apply this technique to reanalyse state-of-the-art weak lensing datasets and assess the contribution of small-scale contaminants to the inferred cosmological parameters. We pay particular attention to the apparent tension between the clumping parameter $S_8$ inferred from Large-Scale Structure and CMB measurements. Finally, we apply the nulling technique to synthetic data representing a Stage-IV LSST-like survey to assess the impact of non-linear scales on upcoming datasets. The main conclusions derived from this analysis are:
  \begin{itemize}
    \item The \dnull and \knull techniques are effective at creating rotated lensing kernels with suppress small-scale and low-redshift support. Removing the most contaminated modes effectively removes all contributions below the chosen comoving distance (for \dnull) or Fourier wavenumber (for \knull), as shown in Figure \ref{fig:qi_qw_desy}.
    \item The signal-to-noise ratio of cosmic shear data is inevitably dominated by the contribution of small-scale structures. The total SNR, both for current and future data, is dominated by the mode that is most affected by small scales and low redshifts (see Figure \ref{fig:snr_nmodes}), which are unfortunately the most difficult to accurately model. Thus, reducing the impact of this contribution inevitably results in significant loss in sensitivity and constraining power.
    \item We analyse the data from DES-Y3 (Figure \ref{fig:desy_params_chimin}). When using the full dataset we find results compatible with \cite{2403.13794}, corresponding to a $2.6\sigma$ shift in $S_8$ with respect to \planck\footnote{Note that \cite{2403.13794} report a $1.7\sigma$ shift when including both baryonic effects and IA. Our $2.6\sigma$ shift is found ignoring IA, and we recover the $1.7\sigma$ shift of \cite{2403.13794} when including both effects.}. When only considering the most contaminated mode, we find 60\% broader constraints on $\Om$, while the constraints on $S_8$ are similar to those found with the full dataset. This is because this mode dominates the overall SNR but is not able to recover tomographic information. If instead we remove this most contaminated mode, both $S_8$ and $\Om$ become more loosely constrained. Interestingly, in this case, both with the \dnull and \knull cases, $S_8$ shows an upward shift ($1$-$2\sigma$ with respect to the full dataset case, see Figure \ref{fig:s8_DES-Y3}), making it more compatible with \planck. However, we show, through a frequentist analysis of these results, that this shift is not statistically significant, and cannot be used to robustly conclude that removing small-scale contamination alleviates the existing $S_8$ tension. 
    \item We find qualitatively the same results with \dnull and \knull and with different values of the comoving-distance and wavemode cut; we also find that leaving the IA parameters free, while decreasing the overall constraining power, does not change the conclusions of our analysis.
    \item We assess the impact of small-scale contaminants for future Stage-IV surveys, and the potential of this method to alleviate uncertainties in their modelling, by analysing a synthetic LSST Y10-like cosmic shear data vector. Also in this case we find that constraints on $S_8$ are dominated by the mode most sensitive to low-redshift small-scale structures, with $\sigma(S_8)$ growing by a factor $\sim3$ if this mode is discarded. Interestingly, the constraints on $\Om$ are, in turn, dominated by the four remaining modes, which, we speculate, are better able to exploit growth information. Ignoring the impact of baryonic effects in the modelling of these data results in a $\gtrsim5\sigma$ bias on $S_8$, which can be reduced to the $\lesssim1\sigma$ level by discarding the most-contaminated mode. Thus, the method described here may be used by future surveys to reduce their sensitivity to small-scale modelling uncertainties, albeit at a significant cost in precision. We also studied the impact of IA on our method, which lead to a huge degradation in the constraints found from the most contaminated mode, but only have a small impact on the remaining modes, which are able to exploit tomographic information.
  \end{itemize}
  We expect the nulling techniques proposed in this work to provide a fundamental framework for understanding the impact of small-scale unmodelled physics in weak lensing analyses. While this method can also be directly used to obtain robust cosmological constraints, this comes at the cost of substantially reducing the dataset constraining power. For this reason, accurate modelling non-linear effects remains crucial for near-future surveys.

\acknowledgments
We would like to thank Pedro Ferreira for useful comments and discussions. GP is supported by the Italian Ministry of University and Research (\textsc{mur}), PRIN 2022 `EXSKALIBUR – Euclid-Cross-SKA: Likelihood Inference Building for Universe's Research', Grant No.\ 20222BBYB9, CUP D53D2300252 0006 and would like to thank the `Angelo della Riccia' Fellowship for supporting part of this work. MZ is supported by STFC. CGG and DA are supported by the Beecroft Trust.

We made extensive use of computational resources at the University of Oxford Department of Physics, funded by the John Fell Oxford University Press Research Fund.

For the purpose of Open Access, the authors have applied a CC BY public copyright licence to any Author Accepted Manuscript version arising from this submission.

\ul{Software}:  We made extensive use of the {\tt numpy} \citep{oliphant2006guide, van2011numpy}, {\tt scipy} \citep{2020SciPy-NMeth}, {\tt astropy} \citep{1307.6212, 1801.02634}, {\tt healpy} \citep{Zonca2019}, {\tt matplotlib} \citep{Hunter:2007} and {\tt GetDist} \citep{Lewis:2019xzd} python packages. We also made extensive use of the {\tt HEALPix} \cite{gorski_healpix_2005} package. 

This paper makes use of software developed for the Large Synoptic Survey Telescope and Euclid. We thank the LSST Project for making their code available as free software at \url{http://dm.lsst.org}. 

\ul{Data:} This project used public archival data from the Dark Energy Survey (DES). Funding for the DES Projects has been provided by the U.S. Department of Energy, the U.S. National Science Foundation, the Ministry of Science and Education of Spain, the Science and Technology Facilities Council of the United Kingdom, the Higher Education Funding Council for England, the National Center for Supercomputing Applications at the University of Illinois at Urbana-Champaign, the Kavli Institute of Cosmological Physics at the University of Chicago, the Center for Cosmology and Astro-Particle Physics at the Ohio State University, the Mitchell Institute for Fundamental Physics and Astronomy at Texas A\&M University, Financiadora de Estudos e Projetos, Funda{\c c}{\~a}o Carlos Chagas Filho de Amparo {\`a} Pesquisa do Estado do Rio de Janeiro, Conselho Nacional de Desenvolvimento Cient{\'i}fico e Tecnol{\'o}gico and the Minist{\'e}rio da Ci{\^e}ncia, Tecnologia e Inova{\c c}{\~a}o, the Deutsche Forschungsgemeinschaft, and the Collaborating Institutions in the Dark Energy Survey.
        
  The Collaborating Institutions are Argonne National Laboratory, the University of California at Santa Cruz, the University of Cambridge, Centro de Investigaciones Energ{\'e}ticas, Medioambientales y Tecnol{\'o}gicas-Madrid, the University of Chicago, University College London, the DES-Brazil Consortium, the University of Edinburgh, the Eidgen{\"o}ssische Technische Hochschule (ETH) Z{\"u}rich,  Fermi National Accelerator Laboratory, the University of Illinois at Urbana-Champaign, the Institut de Ci{\`e}ncies de l'Espai (IEEC/CSIC), the Institut de F{\'i}sica d'Altes Energies, Lawrence Berkeley National Laboratory, the Ludwig-Maximilians Universit{\"a}t M{\"u}nchen and the associated Excellence Cluster Universe, the University of Michigan, the National Optical Astronomy Observatory, the University of Nottingham, The Ohio State University, the OzDES Membership Consortium, the University of Pennsylvania, the University of Portsmouth, SLAC National Accelerator Laboratory, Stanford University, the University of Sussex, and Texas A\&M University.

\newpage
\newpage
\bibliography{biblio}{}
\end{document}